\documentclass[amsmath,amssymb,reprint,twocolumn,superscriptaddress, nobibnotes, prb]{revtex4-1}

\usepackage{lipsum} 

\usepackage[utf8]{inputenc}
\usepackage[T1]{fontenc}

\usepackage[english]{babel}

\usepackage{amsmath,amssymb,amsfonts}
\usepackage{amstext, mathrsfs, textcomp}
\usepackage{mathptmx}
\usepackage{bigints}  
\usepackage{nicefrac}
\usepackage{multirow}
\usepackage{dcolumn}
\usepackage{bm}

\usepackage{subfigure}
\usepackage{graphicx}
\usepackage{xcolor}
\usepackage[export]{adjustbox}

\usepackage[colorlinks=true, citecolor=blue, urlcolor=blue, linkcolor=blue]{hyperref}

\usepackage{changes}

\graphicspath{{Figures/}{Figures/dielectric_cuboid_example/}{Figures/local_mode_coupling/}{Figures/sketches/}{Figures/spheroid/}}

\newcommand{\TITLE}{Generalizing the exact multipole expansion: Density of multipole modes in complex photonic nanostructures}

\newcommand{\checkref}[1]{}

\newcommand{\przero}{\mathbf{p}(\mathbf{r}_0, \omega)}
\newcommand{\pfirstorderrzero}{\mathbf{p}_0(\mathbf{r}_0, \omega)}
\newcommand{\ptoroidalrzero}{\mathbf{p}_t(\mathbf{r}_0, \omega)}
\newcommand{\mrzero}{\mathbf{m}(\mathbf{r}_0, \omega)}
\newcommand{\Prp}{\mathbf{P}(\mathbf{r}', \omega)}

\newcommand{\Einrpp}{\mathbf{E}_0(\mathbf{r}'', \omega)}
\newcommand{\Krprpp}{\overline{\overline{\mathbf{K}}}(\mathbf{r}', \mathbf{r}'', \omega)}
\newcommand{\drp}{\text{d}\mathbf{r}'}
\newcommand{\drpp}{\text{d}\mathbf{r}''}

\newcommand{\alphai}{V_{c}^2\, \overline{\overline{\boldsymbol{\chi}}}_i}

\newcommand{\Kij}{\overline{\overline{\mathbf{K}}}_{ij}}
\newcommand{\Einj}{\mathbf{E}_{0j}}

\begin{document}
	\title{\TITLE}
	\author{\firstname{Cl\'ement} \surname{Majorel}}
	\affiliation{CEMES-CNRS, Universit\'e de Toulouse, CNRS, UPS, 31000 Toulouse, France}
	
	\author{\firstname{Adelin} \surname{Patoux}}
	\affiliation{CEMES-CNRS, Universit\'e de Toulouse, CNRS, UPS, 31000 Toulouse, France}
	\affiliation{LAAS-CNRS, Universit\'e de Toulouse, 31000 Toulouse, France}
	\affiliation{AIRBUS DEFENCE AND SPACE SAS, 31000 Toulouse, France}
	
	\author{\firstname{Ana} \surname{Estrada-Real}}
	\affiliation{LAAS-CNRS, Universit\'e de Toulouse, 31000 Toulouse, France}
	\affiliation{INSA-CNRS-UPS, LPCNO, Universit\'e de Toulouse, 31000 Toulouse, France}
	
	\author{\firstname{Bernhard} \surname{Urbaszek}}
	\affiliation{INSA-CNRS-UPS, LPCNO, Universit\'e de Toulouse, 31000 Toulouse, France}
	
	\author{\firstname{Christian} \surname{Girard}}
	\affiliation{CEMES-CNRS, Universit\'e de Toulouse, CNRS, UPS, 31000 Toulouse, France}
	
	\author{\firstname{Arnaud} \surname{Arbouet}}
	\affiliation{CEMES-CNRS, Universit\'e de Toulouse, CNRS, UPS, 31000 Toulouse, France}
	
	\author{\firstname{Peter R.} \surname{Wiecha}}
	\email[e-mail~: ]{pwiecha@laas.fr}
	\affiliation{LAAS-CNRS, Universit\'e de Toulouse, 31000 Toulouse, France}

	\begin{abstract}
		The multipole expansion of a nano-photonic structure's electromagnetic response is a versatile tool to interpret optical effects in nano-optics, but it only gives access to the modes that are excited by a specific illumination. In particular the study of various illuminations requires multiple, costly numerical simulations.
		Here we present a formalism we call ``generalized polarizabilities'', in which we combine the recently developed exact multipole decomposition [Alaee et al., Opt. Comms. 407, 17-21 (2018)] with the concept of a generalized field propagator.
		After an initial computation step, our approach allows to instantaneously obtain the exact multipole decomposition for any illumination.
		Most importantly, since all possible illuminations are included in the generalized polarizabilities, our formalism allows to calculate the total density of multipole modes, regardless of a specific illumination, which is not possible with the conventional multipole expansion. 
		Finally, our approach directly provides the optimum illumination field distributions that maximally couple to specific multipole modes.
%
%
%
%
%
%
		The formalism will be very useful for various applications in nano-optics like illumination-field engineering, or meta-atom design e.g. for Huygens metasurfaces. 
		We provide a numerical open source implementation compatible with the \textit{pyGDM} python package.
	\end{abstract}

	\maketitle
	\section{Introduction}
	
	Studying the interaction of light with structures of sizes smaller or similar to the wavelength has tremendous importance for various scientific areas and related applications.
	Already the broad area of nano-optics covers research on a vast range of phenomena such as resonant or directional scattering,\cite{kuznetsovOpticallyResonantDielectric2016, fuDirectionalVisibleLight2013, wiechaStronglyDirectionalScattering2017} polarization conversion,\cite{katsGiantBirefringenceOptical2012} nonlinear scattering of e.g. second or third harmonic light,\cite{rodrigoCoherentControlLight2013, shcherbakovEnhancedThirdHarmonicGeneration2014, wiechaEnhancedNonlinearOptical2015} optical forces\cite{girardTheoreticalAnalysisLightinductive1994, chaumetCoupledDipoleMethod2000} or nano-scale heat generation.\cite{baffouThermoplasmonicsUsingMetallic2013, girardDesigningThermoplasmonicProperties2018}
	These effects are used for instance to study atmospheric or astrophysical particles, \cite{mulhollandLightScatteringAgglomerates1994, huntemannDiscreteDipoleApproximation2011, draineDiscreteDipoleApproximationIts1988} and they have many practical applications, for instance in medicine for hyperthermia treatments or rapid antigen tests.\cite{cherukuriTargetedHyperthermiaUsing2010, stockmanNanoplasmonicsPhysicsApplications2011}
	Understanding and modeling of the interaction of nanostructures with light is also essential for optical metasurfaces.\cite{genevetRecentAdvancesPlanar2017}
	
	An important tool in the description and  interpretation of nano-scale light-matter interaction is the modal analysis of the optical response. 
	A powerful method is the quasinormal mode (QNM) expansion, aiming at the identification of all available resonant modes of an open system such as a photonic nanostructure.\cite{baiEfficientIntuitiveMethod2013, lalanneLightInteractionPhotonic2018}
	However, QNM expansions are often not straightforward, in particular the normalization of QNMs is a difficult task, due to the description of these modes using complex eigenfrequencies.\cite{kristensenNormalizationQuasinormalModes2015, chenGeneralizingNormalMode2019}
	
	A somewhat simpler, yet very useful modal analysis of an already excited nano-photonic system is a multipole expansion of its induced polarization density.\cite{jacksonClassicalElectrodynamics1999}
	The conventional expansion can be found in any electro-dynamics textbook,\cite{jacksonClassicalElectrodynamics1999} and is based on a long-wavelength approximation for the fields emitted by the multipole moments. 
	Recently, exact expressions for the multipole expansion beyond the long-wavelength limit have been derived, that yield accurate results also in the case of larger nanostructures.\cite{alaeeElectromagneticMultipoleExpansion2018, alaeeExactMultipolarDecompositions2019, evlyukhinMultipoleDecompositionsDirectional2019}
	Naturally, for an increasing structure size the number of required multipole terms rapidly increments, which is why the method is best suited for structures of sizes not much larger than the wavelength.
	While in plasmonic nanostructures usually the electric dipole dominates,\cite{arangoPolarizabilityTensorRetrieval2013}
	resonant dielectric nanostructures often possess higher order modes due to retardation effects.
	Finally, dielectric structures confine light less efficiently than plasmonic particles, therefore their multipolar modes occur typically at sizes where the long-wavelength multipole expansion is no longer accurate.\cite{kuznetsovOpticallyResonantDielectric2016}
	Consequently, the exact multipole expansion is of particular relevance for dielectric nano-resonantors.


	From a technical point of view, the application of the multipole expansion to photonic nanostructures is straightforward, and can be done in combination with any numerical solver.\cite{evlyukhinMultipoleLightScattering2011, hinamotoMENPOpensourceMATLAB2021, munDescribingMetaAtomsUsing2020}
	But, in contrast to the QNM analysis which provides information about the nanostructure itself, for the multipole expansion the mode-basis is chosen before and is then used for the analysis of the electric polarization density inside the nanostructure upon illumination. 
	It thus offers merely an analysis of an excited state, but no rigorous description of fundamental, resonant properties of the nano-resonator.
	This means that an illumination needs to be chosen a priori, and a new simulation needs to be performed for every change in the illumination.

	To develop a more general modal analysis tool in the multipole basis, here we build on the exact multipole expansion of the polarization density,\cite{alaeeElectromagneticMultipoleExpansion2018} and combine it with the concept of a generalized field propagator.\cite{martinGeneralizedFieldPropagator1995}
	This allows to obtain a set of generalized polarizability tensors for each multipole order, establishing a direct link between an arbitrary illumination field and the induced modes in the exact multipole expansion. 
	In contrast to the classical multipole expansion, the generalized polarizabilities allow to calculate the total mode density for the different multipoles, regardless of a specific illumination.
	They can be used to study and visualize local properties of light-matter interaction inside a nanostructure, and, as a by-product, they provide the optimum illumination field distribution for maximum coupling to the respective multipole modes.
	Finally, once calculated, the generalized polarizabilities are a computationally very cheap approach to obtain a model for the optical response under arbitrary illuminations and they can be stored efficiently thanks to their light memory footprint.
	We discuss our formalism in comparison with the very popular and accurate T-matrix method. 
	
	We demonstrate the potential of our formalism by analyzing the available modes in a dielectric nano-scatterer and their selective excitation under various illuminations.
	We furthermore study a dielectric Huygens source and find that a plane wave couples to different multipole modes depending on the angle of incidence, which is a main reason for the limited efficiencies of dielectric Huygens metasurfaces.

	\section{Formalism}

	When studying the optical interaction of a nanostructure with an external illumination, it is usually insightful to get an approximate, but physically meaningful model for the nanostructure's optical response.
	To this end, a multipole expansion of the electric polarization density inside a nanostructure can be performed.\cite{jacksonClassicalElectrodynamics1999, alaeeElectromagneticMultipoleExpansion2018, evlyukhinMultipoleLightScattering2011}
	This gives access to the effective electric and magnetic dipole modes, quadrupole modes, etc..., that are induced by the interaction of the nanostructure with an external illumination.\cite{sersicMagnetoelectricPointScattering2011, arangoPolarizabilityTensorRetrieval2013, kuznetsovOpticallyResonantDielectric2016, wuIntrinsicMultipolarContents2020}
	In fact, considering just the dipolar modes of a nano-structure can already give a quite accurate picture of the physics at play, especially in the far-field to which higher order modes like quadrupoles usually couple more inefficiently. 
	However higher order contributions readily lead to localized phenomena in the near-field and, in case of high quality factors, can still couple to the far-field in a significant manner.\cite{lunnemannLocalDensityOptical2016, patouxPolarizabilitiesComplexIndividual2020, abujetasCoupledElectricMagnetic2020}
	
	\begin{figure}[t]
		\centering
		\includegraphics[width=\columnwidth]{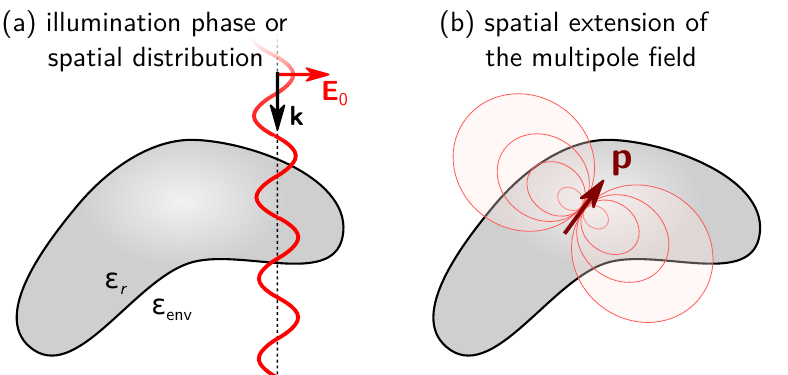}
		\caption{
			The long wavelength approximation describes structures which are very small compared to the wavelength. If the structures become larger the phase and field distribution over its volume cannot be assumed constant any longer. Retardation effects influence the multipole expansion at two levels: 
			(a) The true spatial variations of the illumination field and its phase across the nanostructure and (b) the spatial distribution of the field emitted by the multipoles in the expansion.
		}
		\label{fig:scheme_phase_effects}
	\end{figure}

	In the case of atoms, molecules or very small nanostructures, the illumination field can be considered constant at the scale of the nanoparticle. 
	This also allows to describe the structure with polarizability tensors, relating the illumination field $\mathbf{E}_0(\mathbf{r}_0, \omega)$ at the particle's position ($\mathbf{r}_0$) to an induced multipole moment.\cite{buckinghamPermanentInducedMolecular1967, arangoPolarizabilityTensorRetrieval2013, munDescribingMetaAtomsUsing2020}
	However, if a nanostructure is larger (illustrated in figure~\ref{fig:scheme_phase_effects}a), the quasistatic approximation does not hold any longer.
	An inhomogeneous illumination field can even lead to entirely new effects. An example are magnetic resonances in dielectric nanostructures.
	The latter are a result of optical vortices, that are created by a varying phase of the illumination along the nanostructure's extension.\cite{kuznetsovOpticallyResonantDielectric2016, baranovModifyingMagneticDipole2017, wiechaDecayRateMagnetic2018, patouxPolarizabilitiesComplexIndividual2020}
	For larger nanostructures we therefore first need to solve rigorously the light-matter interaction, before we expand the induced electric polarization density inside the particle into multipole moments.
	
	The electric polarization density $\mathbf{P}(\mathbf{r}, \omega)$ inside nanoparticles of arbitrary shape can be obtained only numerically. 
	Here we will use the Green's Dyadic Method (GDM), a frequency domain, volume integral approach.\cite{girardFieldsNanostructures2005}
	In the GDM, we start with the Lippmann-Schwinger equation (cgs units)
	\begin{multline}\label{eq:lippmann_schwinger}
		\mathbf{E}(\mathbf{r}', \omega) = \mathbf{E}_0(\mathbf{r}', \omega) + \\ \int\limits_{V_{ns}} \drpp\ \overline{\overline{\mathbf{G}}}(\mathbf{r}', \mathbf{r}'', \omega) \cdot \overline{\overline{\boldsymbol{\chi}}}(\mathbf{r}'', \omega) \cdot \mathbf{E}(\mathbf{r}'', \omega) \, ,
	\end{multline}
	that relates the induced electric field $\mathbf{E}(\mathbf{r}', \omega)$ and the unperturbed illumination field $\mathbf{E}_0(\mathbf{r}', \omega)$ in a self-consistent way.
	$\overline{\overline{\boldsymbol{\chi}}}$ is the electric susceptibility tensor of the nanostructure, corresponding to the difference between the relative permittivities of nanostructure and environment $\overline{\overline{\boldsymbol{\chi}}} = (\boldsymbol{\epsilon}_r-\boldsymbol{\epsilon}_{\text{env}})/4\pi$.
	$\overline{\overline{\mathbf{G}}}$ is the dyadic Green's function of the bare environment.
	The integral runs over the entire volume $V_{ns}$ occupied by the nanostructure.
	From this it is possible to derive a so-called \textit{generalized field propagator} $\Krprpp$, that directly relates the incident electric field $\mathbf{E}_0(\mathbf{r}'', \omega)$ to the induced local electric polarization $\mathbf{P}(\mathbf{r}', \omega)$:\cite{martinGeneralizedFieldPropagator1995}
	\begin{equation}\label{eq:generalized_propa}
		\Prp = \overline{\overline{\boldsymbol{\chi}}}(\mathbf{r}', \omega) \cdot \int\limits_{V_{ns}} \drpp\ \Krprpp \cdot \Einrpp \, .
	\end{equation}
	The derivation of $\Krprpp$ for locations inside the structure is sketched in appendix~\ref{appendix:generalized_propagator}.
	Note that both locations $\mathbf{r}'$ and $\mathbf{r}''$ are inside the nanostructure and that while in our notation we use the electric polarization density, it is equivalent to using the current density, since for time-harmonic fields $\mathbf{P}(\omega) = -\mathbf{j}(\omega)/\text{i}\omega$.\cite{novotnyPrinciplesNanooptics2006}
	
	The generalized propagator and thus the electric polarization $\mathbf{P}$ at any location $\mathbf{r}'$ inside a nanostructure can now be obtained numerically by discretizing the nanostructure on a regular grid into a number of $N_c$ mesh-cells, each of volume $V_c$. 
	Such discretization allows to numerically solve the optical Lippmann-Schwinger equation with conventional inversion techniques, giving access to the generalized propagator. 
	For details, we refer to related literature.\cite{girardFieldsNanostructures2005, girardShapingManipulationLight2008, wiechaPyGDMPythonToolkit2018, wiechaPyGDMNewFunctionalities2022}
	Note that the discretization also transforms the integral in equation~\eqref{eq:generalized_propa} into a finite sum over the nanostructure mesh-cells:
	\begin{equation}\label{eq:generalized_propa_discretized}
		\begin{aligned}
		\mathbf{P}(\mathbf{r}_i, \omega) & = \overline{\overline{\boldsymbol{\chi}}}(\mathbf{r}_i, \omega) \cdot \sum\limits_{j=1}^{N_c} V_c\ \overline{\overline{\mathbf{K}}}(\mathbf{r}_i, \mathbf{r}_j, \omega) \cdot \mathbf{E}_0(\mathbf{r}_j, \omega) \\
		& = V_{c}\, \overline{\overline{\boldsymbol{\chi}}}_i \cdot \sum\limits_{j}^{N_c} \Kij\cdot \Einj \, .
		\end{aligned}
	\end{equation}
	In the second line of equation~\eqref{eq:generalized_propa_discretized} we introduced an abbreviated notation, where indices $i$ and $j$ indicate evaluation at, respectively, the $i$th and $j$th mesh-cell in the discretization. For the sake of readability we also omit the dependence on the frequency $\omega$.
	We will use this notation in the following for all discretized equations.

	Now we have access to the distribution of the electric polarization density inside an arbitrary nanostructure, which we can subsequently expand into a series of multipole contributions.
	As mentioned above the conventional multipole expansion\cite{jacksonClassicalElectrodynamics1999} is based on a long-wavelength approximation, and valid only if the field corresponding to the multipole moments  (dipole, quadrupole, etc...) can be described in this approximation over the entire nanostructure volume (see figure~\ref{fig:scheme_phase_effects}b). 
	When the size of the nanostructure increases and begins to be comparable to the optical wavelength in its material, the conventional equations for the multipole expansion are increasingly inaccurate. 
	For size parameters $a/\lambda \gtrsim 0.5$ (with $a$ being the diameter or total length of the structure), this long-wavelength multipole expansion becomes essentially invalid.\cite{alaeeElectromagneticMultipoleExpansion2018}

	In Refs.~\onlinecite{alaeeElectromagneticMultipoleExpansion2018, alaeeExactMultipolarDecompositions2019} Alaee et al. have shown recently, that \textit{exact} equations for the multipole moments can be derived, valid for any particle size, and not significantly more complicated than their long-wavelength counterparts. 
	The exact electric and magnetic dipole moments $\mathbf{p}$ and $\mathbf{m}$, induced in a nanostructure are found to be:
	\begin{subequations}\label{eq:multipoles_p}
		\begin{align}
			\przero &= \pfirstorderrzero\ +\ \ptoroidalrzero \\
			\pfirstorderrzero &= \int\limits_{V_{ns}} \drp\ \Prp\ j_0(k r') \\
			\ptoroidalrzero &=  \frac{k^2}{2} \int\limits_{V_{ns}} \drp\ 
			\bigg[3\big(\mathbf{r}'\cdot \Prp\big) \mathbf{r}' \\
			& \quad\quad\quad\quad\quad - r'^2\, \Prp \bigg] \frac{j_2 (kr')}{(kr')^2}\nonumber
		\end{align}
	\end{subequations}

	\begin{equation}\label{eq:multipoles_m}
	\mrzero = \frac{-3\text{i}k}{2} \int\limits_{V_{ns}} \drp \Big(\mathbf{r}' \times \Prp \Big) \frac{j_1(k r')}{kr'}
	\end{equation}
	$k=2\pi n_{\text{env}}/\lambda_0$ is the wavenumber in the surrounding medium (of refractive index $n_{\text{env}}$)  and $j_n$ is the $n$th order spherical Bessel function of the first kind.
	The electric dipole moment consists of two contributions. 
	Its first order term $\mathbf{p}_0$ and higher order contributions described by a second term $\mathbf{p}_t$, often called the ``toroidal'' dipole  moment.\cite{dubovikToroidMomentsElectrodynamics1990}
	While the toroidal multipoles are no orthogonal modes by themselves,\cite{alaeeElectromagneticMultipoleExpansion2018} we will develop both contributions separately in order to be able to distinguish them numerically.
	$\mathbf{r}_0$ is the expansion location of the multipole series. 
	For convenience $\mathbf{r}_0$ is at the origin of our coordinate system. 
	We use the nanostructure's center of gravity as location for the series expansion.\cite{evlyukhinMultipoleLightScattering2011}

	By applying the above described volume discretization to equation~\eqref{eq:multipoles_p} and combining it with equation~\eqref{eq:generalized_propa_discretized}, we obtain for the electric dipole moments: 
	\begin{subequations}\label{eq:electric_dipole_gen_propagator}
		\begin{align}
		\mathbf{p}_{0}(\mathbf{r}_0, \omega) &= \sum\limits_{i}^{N_c} \Big( \alphai\cdot \sum\limits_{j}^{N_c} \Kij\cdot \Einj \Big)\ j_0(k r_i) \\
		\mathbf{p}_t(\mathbf{r}_0, \omega) &= \frac{k^2}{2} \sum\limits_{i}^{N_c} 
			\Bigg[ 
		       3\bigg(\mathbf{r}_i\cdot \Big(\alphai \cdot\sum\limits_{j}^{N_c} \Kij\cdot \Einj \Big)\ \bigg) \mathbf{r}_{i} \\
		       & \quad\quad - r_i^2\ \bigg( \alphai \cdot \sum\limits_{j}^{N_c} \Kij\cdot \Einj\bigg)\
  		    \Bigg]
		       \frac{j_2 (kr_i)}{(kr_i)^2} \, ,
		       \nonumber
		\end{align}
	\end{subequations}
	where ``\,$\cdot$\,'' is the dot product between two tensors, or the scalar product between two vectors.

	The core idea of this work is to interchange the summation order of indices $i$ and $j$  and to move the illumination field at each meshcell $\mathbf{E}_{0j}$ out of the sum over index $i$. 
	This will allow us to evaluate the sum over $i$ without prior knowledge of the illumination.
	In Eqs.~\eqref{eq:electric_dipole_gen_propagator} this is straightforward for all terms except for the first term of the toroidal dipole $\mathbf{p}_t$, which is a scalar product \textit{after} multiplication of $\mathbf{K}_{ij}$ with the illumination field $\mathbf{E}_{0j}$. 
	In this case we need to introduce a further sum over the three Cartesian vector components ($x, y, z$) of the toroidal dipole moment, in order to perform the scalar product after the evaluation of the light matter interaction. 
	We get for the $a$-component of the dipole moment vectors
	\begin{subequations}\label{eq:electric_dipole_gen_propagator_interchanged_summation}
	\begin{align}
		\label{eq:electric_dipole_gen_propagator_interchanged_summation_a}
		p_{0}^{a}(\mathbf{r}_0, \omega) &= V_{c}^2 \sum\limits_{j}^{N_c} 
				\Bigg[ \sum\limits_{i}^{N_c} \Big( \chi^{a}_{i,\gamma} K^{\gamma}_{ij,\epsilon} \Big)\ j_0(k r_i) \Bigg] E_{0j}^{\epsilon}\\
		\label{eq:electric_dipole_gen_propagator_interchanged_summation_b}
		p_{t}^{a}(\mathbf{r}_0, \omega) &= \frac{k^2 V_c^2}{2}  \sum\limits_{l=1}^3 \sum\limits_{j}^{N_c}\ 
		\Bigg[ \sum\limits_{i}^{N_c} 
		\Bigg( 3\Big(r_{i,l} \chi_{i,\gamma}^{l} K_{ij,\epsilon}^{\gamma} \Big) r_{i}^{a} \\
		& \quad\quad - \frac{r_i^2}{3}\ \Big( \chi_{i,\gamma}^{a} K_{ij,\epsilon}^{\gamma} \Big) \Bigg)
		\frac{j_2 (kr_i)}{(kr_i)^2}
		\Bigg] E_{0j}^{\epsilon} \, .
		\nonumber
	\end{align}
	\end{subequations}
	Following Einstein's convention, tensors are contracted over Greek lower case letter indices that occur twice within a product. 
	The factor $1/3$ in the second term of Eq.~\eqref{eq:electric_dipole_gen_propagator_interchanged_summation_b} comes from the conversion of the scalar product $\mathbf{r}_i \cdot \mathbf{P}_i$ into a sum over the three Cartesian coordinates $l$, that has been explicitly moved out of the sum over $i$ and affects now the entire term in square brackets.
	The terms that act on the incident electric field are thus $N_c$ rank 2 tensors for $\mathbf{p}_0$, and $N_c$ rank 3 tensors for~$\mathbf{p}_t$:
	\begin{subequations}\label{eq:electric_dipole_gen_polarizabilities}
		\begin{align}
			\alpha^{p_0, a}_{j,\epsilon}(\mathbf{r}_0, \omega) &= V_{c}^2 
			\sum\limits_{i}^{N_c} \Big(\chi^{a}_{i,\gamma} K^{\gamma}_{ij,\epsilon} \Big)\ j_0(k r_i)\\
			\alpha^{p_t, a}_{j,l\epsilon}(\mathbf{r}_0, \omega) &= \frac{k^2 V_{c}^2 }{2} 
			\sum\limits_{i}^{N_c} 
			\Bigg( 3\Big(r_{i,l} \chi_{i,\gamma}^{l} K_{ij,\epsilon}^{\gamma} \Big) r_{i}^{a} \\
			& \quad\quad - \frac{r_i^2}{3}\ \Big( \chi_{i,\gamma}^{a} K_{ij,\epsilon}^{\gamma} \Big) \Bigg)
			\frac{j_2 (kr_i)}{(kr_i)^2} \, .
			\nonumber
		\end{align}
	\end{subequations}
	Each tensor $\overline{\overline{\alpha}}^{p_0}_{j}$ and $\overline{\overline{\alpha}}^{p_t}_{j}$ describes the contribution of the $j$th mesh-cell to, respectively, the electric dipole moment and the toroidal dipole moment in the multipole expansion:
	\begin{equation}
		\mathbf{p}(\mathbf{r}_0, \omega) = 
		  \overbrace{\sum\limits_{j}^{N_c} \overline{\overline{\alpha}}^{p_0}_{j} \cdot \mathbf{E}_{0j}}^{\text{dipole}}\ +\ 
		  \overbrace{\sum\limits_{l=1}^3 \sum\limits_{j}^{N_c} \overline{\overline{\alpha}}^{p_t}_{j,l} \cdot \mathbf{E}_{0j}}^{\text{toroidal dipole}} \, .
	\end{equation}
	We call these the \textit{generalized electric-electric dipole polarizability} tensors. Generalized, since they allow to obtain the effective dipole moment induced in a nanostructure by an arbitrary illumination field. 
	In other words, $\overline{\overline{\alpha}}^{p_0}_{j}$ describes the strength of light-matter interaction at the location $\mathbf{r}_j$ in the nanostructure, for inducing an effective electric dipole moment. 
	Likewise, $\overline{\overline{\alpha}}^{p_t}_{j}$ describes the local coupling strength of an illumination field at $\mathbf{r}_j$, to the toroidal dipole moment.

	\begin{figure}[t]
	\centering
	\includegraphics[width=\columnwidth]{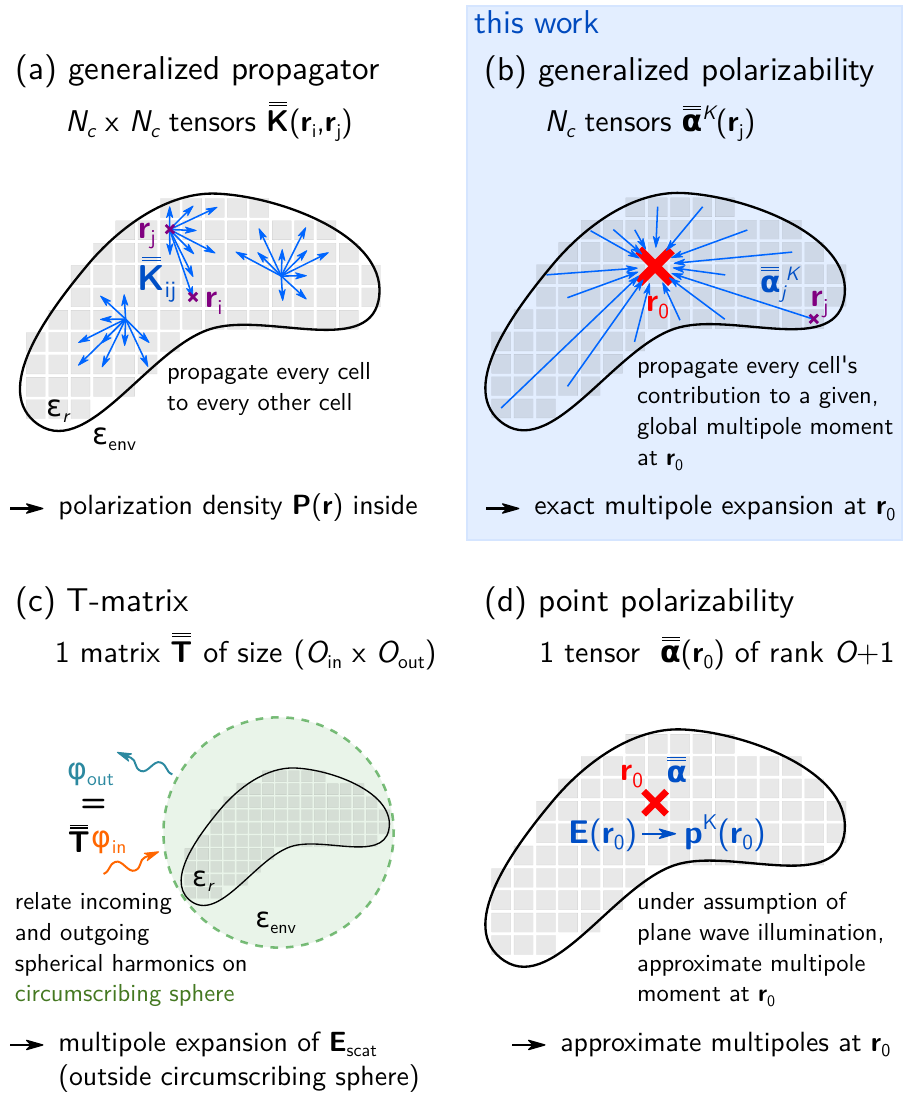}
	\caption{
		A particle of relative permittivity $\epsilon_r$ is placed in an environement of permittivity $\epsilon_{\text{env}}$, and its volume is discretized on a regular mesh.
		(a) The generalized field propagator $\overline{\overline{\mathbf{K}}}$ establishes a relation between the illumination field inside the structure and the spatial distribution of the electric polarization by propagating the zeroth order dipole moment of each meshcell to the locations of all other meshcells. 
		(b) The generalized polarizabilities $\overline{\overline{\boldsymbol{\alpha}}}^K$, presented in this work, propagate the zeroth order dipole moments of all meshcells to the location $\mathbf{r}_0$ of the multipole expansion. 
		It thus represents each meshcell's contribution to the total multipole moment. The superscript $K$ indicates the type of multipole moment (e.g. electric or magnetic dipole or quadrupole).
		For illustration, only a few propagation vectors are shown.
		(c) The T-matrix establishes a relation between expansion coefficients of incoming and outgoing (scattered) spherical vector harmonics (denoted as $\varphi_{\text{in}}$, respectively $\varphi_{\text{out}}$). Fields are valid outside the circumscribing sphere.
		(d) Point polarizability models approximate the multipole moment ($\mathbf{p}^K$) under the assumption of plane wave illumination.
	}
	\label{fig:gen_propagator_gen_pola}
\end{figure}

	Proceeding in the same way with equation~\eqref{eq:multipoles_m}, we obtain the \textit{generalized electric-magnetic polarizabilities}:
	\begin{equation}\label{eq:magnetic_dipole_gen_propagator_interchanged_summation}
		m^{a}(\mathbf{r}_0, \omega) = \frac{-3\text{i}k V_c^2}{2} \sum\limits_{j}^{N_c} \Bigg[ \sum\limits_{i}^{N_c} \Big( \epsilon^a_{\zeta\kappa}\ r_i^{\zeta} \chi_{i,\gamma}^{\kappa} K_{ij,\epsilon}^{\gamma}\Big)  \frac{j_1(k r_i)}{kr_i}\Bigg] E_{0,j}^{\epsilon}
	\end{equation}
	\begin{equation}\label{eq:magnetic_dipole_gen_polarizabilities}
		\alpha^{m, a}_{j,\epsilon}(\mathbf{r}_0, \omega) = \frac{-3\text{i}k V_c^2}{2} \sum\limits_{i}^{N_c} \Big( \epsilon^a_{\zeta\kappa}\ r_i^{\zeta} \chi_{i,\gamma}^{\kappa} K_{ij,\epsilon}^{\gamma}\Big)  \frac{j_1(k r_i)}{kr_i}
	\end{equation}
	where $\epsilon^a_{\zeta\kappa}$ is the Levi-Civita symbol, describing the vector product $\mathbf{r}_i \times \mathbf{P}_i $.
	The tensors $\overline{\overline{\alpha}}^{m}_j$ link the illumination electric field distribution $\mathbf{E}_0$ inside the structure to the induced, total magnetic dipole moment $\mathbf{m}$:
	%
	\begin{equation}
		\mathbf{m}(\mathbf{r}_0, \omega) = \sum\limits_{j}^{N_c} \overline{\overline{\alpha}}^{m}_{j} \cdot \mathbf{E}_{0j}
	\end{equation}
	This can be done in the same way also for the higher order multipoles. The expressions for the generalized polarizabilities  of electric quadrupole and magnetic quadrupole are given in the Appendix~\ref{appendix:quadrupole_terms}.

	In summary, with the generalized polarizability tensors we now have a tool that allows
	to directly obtain the multipole expansion for arbitrary distributions of the illumination field $\mathbf{E}_0$ on a fixed nanostructure geometry, without requiring any further numerical simulation. For each multipole we only need to perform $N_c$ matrix-vector multiplications of the generalized polarizability tensors with the illumination field.
	In case of the generalized propagator on the other hand, we require a total of $N_c^2$ matrix vector multiplications. In return we get the full electric polarization density inside the structure, while the generalized polarizabilities only give access to its multipole expansion. This is illustrated in figure~\ref{fig:gen_propagator_gen_pola}a-b.
	
	Besides the faster evaluation, the fact that for a structure discretized in $N_c$ mesh-cells, we require only $N_c$ generalized polarizability tensors, instead of $N_c^2$ generalized field propagator tensors $\Kij$ has important implications on the memory requirements.
	Let's illustrate the scaling with a structure of 10,000 mesh-cells. 
	Storage of the electric and magnetic dipolar response with single precision floating point numbers requires only 1.72\,MBytes (this is including the toroidal dipole). The electric and magnetic quadrupole moments add another 4.12\,MBytes.
	On the other hand, 3.43\,GBytes are required to store the generalized propagators for the same structure.
	Creating an extensive database of the generalized polarizabilities is obviously more realistic than saving the generalized field propagators for a larger set of nanostructures.

	\subsection*{T-matrix method and point polarizabilities}

	Before we demonstrate the capabilities of the generalized polarizabilities, we want to briefly position the formalism with respect to conceptually related methods.
	
	The \textit{T-matrix method} (TMM) is also based on an expansion of the optical response in vector spherical harmonics, it can be seen as a generalization of Mie theory.\cite{watermanMatrixFormulationElectromagnetic1965,mishchenkoScatteringAbsorptionEmission2002, mishchenkoMultipleScatteringRadiative2008, litvinovDerivationExtendedBoundary2008}
	The T-matrix contains the field expansion coefficients that relate the incoming with the outgoing fields. These are commonly obtained by point matching on a sphere that encloses the nano-scatterer, as illustrated in figure~\ref{fig:gen_propagator_gen_pola}c. The T-matrix for a spherical particle is hence of diagonal form, containing exactly the Mie coefficients.
	In the T-matrix obtained via the conventional point-matching method,\cite{lokeTmatrixCalculationDiscrete2009, fruhnertComputingTmatrixScattering2017, bertrandGlobalPolarizabilityMatrix2020} the fields are valid only outside the circumscribing sphere. Efforts to extend the TMM validity usually come at the cost of other limitations like a significant increase of computational complexity or a reduced accuracy.\cite{egelLightScatteringOblate2016, demesyScatteringMatrixArbitrarily2018, martinTmatrixMethodClosely2019}
	
	The greatest strength of the TMM is the possibility to couple large numbers of nanostructures and calculate multi-scattering in complex systems with very good accuracy.\cite{mishchenkoMultipleScatteringRadiative2008, pattelliRolePackingDensity2018, werdehausenModelingOpticalMaterials2020, skardaLowoverheadDistributionStrategy2022}
	While periodicities can be implemented in the Green's tensor,\cite{abujetasCoupledElectricMagnetic2020, rahimzadeganComprehensiveMultipolarTheory2022} and scattering between few scatterers would be in principle possible, describing many coupled structures of different shape with the generalized polarizabilities would rapidly lead to huge systems of coupled equations, as a result of the spatial discretization of the illumination. 
	In consequence, while the approach is ideal for the analysis of single (possibly periodic) nano-scatterers, the TMM is clearly the method of choice for multi-scattering simulations of complex arrangements.

	Another popular concept is the \textit{point polarizability}.
	As illustrated in figure~\ref{fig:gen_propagator_gen_pola}d, it is defined as the tensorial proportionality factor, linking the field at the location $\mathbf{r}_0$ of the multipole expansion to the multipole moment induced by a plane wave.\cite{sersicMagnetoelectricPointScattering2011, arangoPolarizabilityTensorRetrieval2013, patouxPolarizabilitiesComplexIndividual2020, rahimzadeganComprehensiveMultipolarTheory2022}
	Since the multipoles of the point polarizabilities and the T-matrix use the same expansion basis (vector spherical harmonics), higher order point-polarizability tensors of a structure can be derived from the T-matrix (and vice-versa), essentially via a coordinate transformation.\cite{munDescribingMetaAtomsUsing2020}

	Our generalized polarizability tensors relate the illumination field at each position inside the nano-scatterer to a Cartesian multipole moment. Hence, in contrast to the above mentioned techniques, neither does the illumination need to be expanded in spherical harmonics (TMM), nor approximated as a plane wave (point-polarizabilities). 
	In consequence, and in the limit of using only the first few expansion terms, our approach promises a better accuracy inside the TMM circumscribing sphere (see also SI figure~S6), and offers highest accuracy for the description of complex spatial distributions of non-plane wave illuminations (see also SI figure~S7).
	
	Finally, our approach requires a single computational invest to calculate the set of generalized polarizabilities. Like the T-matrix and the point polarizabilities, once calculated, our method is very efficient.
	In particular the T-matrix calculation usually requires a series of many simulations, hence its extraction is computationally expensive.\cite{fruhnertComputingTmatrixScattering2017}

	\begin{figure}[t]
		\centering
		\includegraphics[width=\columnwidth]{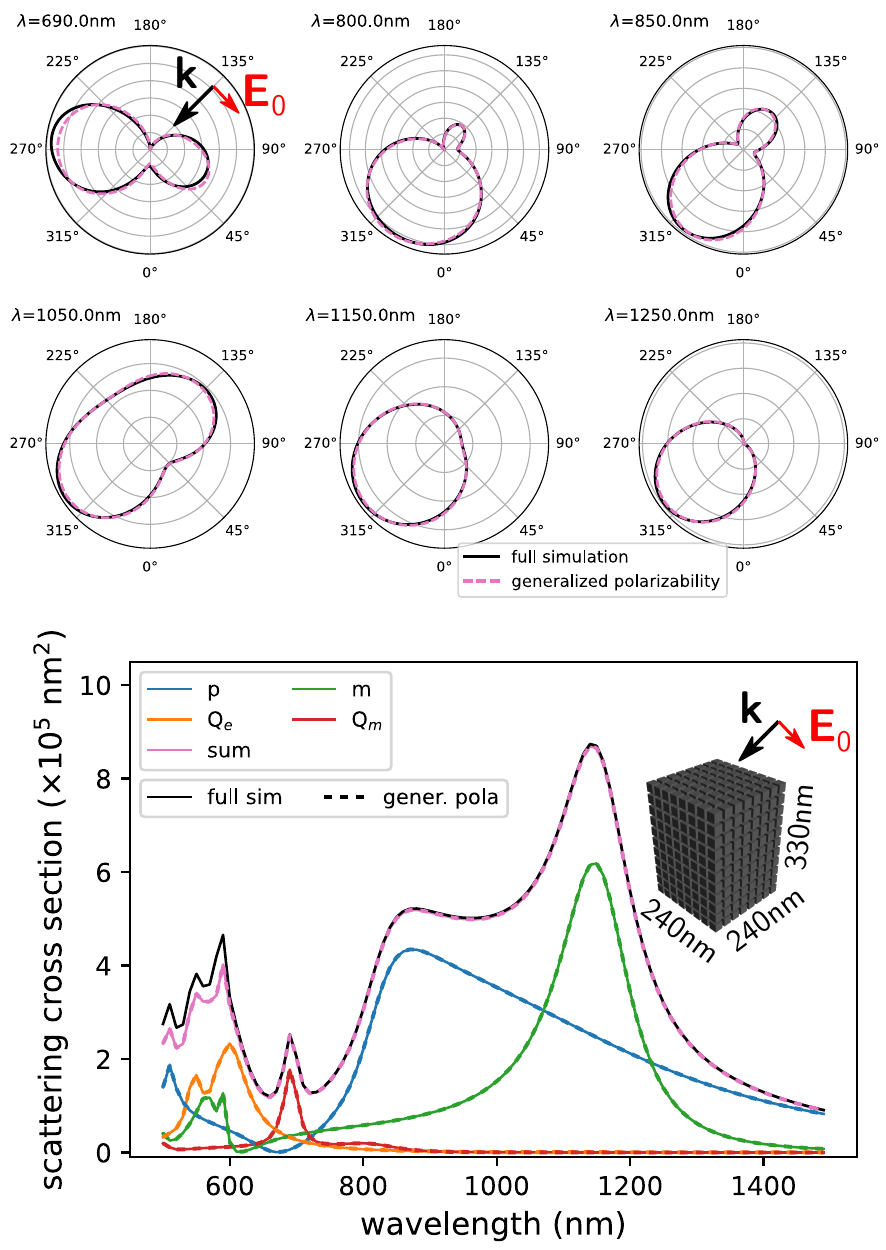}
		\caption{
			Comparison of far-field scattering simulations and analogous calculations with the generalized polarizability tensors. The considered nano-object is a nano-cuboid of dimensions $240\times 240\times 330\,$nm$^3$ made of a lossless material ($n=3.5$), placed in air ($n_{\text{env}}=1$).
			Illumination is a $p$ polarized, oblique plane wave at an incident angle of $135^{\circ}$. 
			Top: radiation patterns in the scattering plane for selected wavelengths. Solid black lines correspond to full field simulations, dashed purple lines to multipoles via the generalized polarizability.
			Bottom: total far-field scattering cross section (black line) and its multipole decomposition from full simulations (solid lines) as well as obtained via the generalized polarizabilities (dashed lines of same color).
		}
		\label{fig:spectra_radpattern}
	\end{figure}

	
	\section{Benchmark}

	\begin{figure*}[t]
		\centering
		\includegraphics[width=\textwidth]{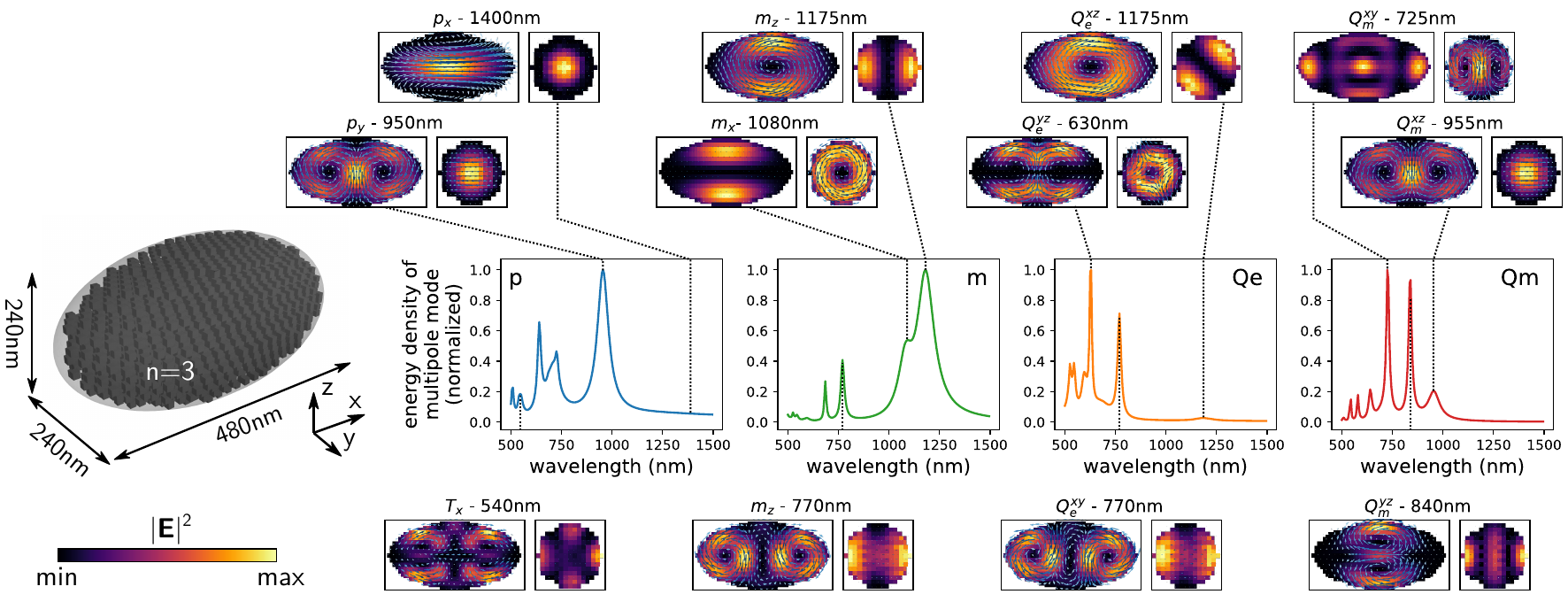}
		\caption{
			Spectra of the normalized total mode energy density for the different multipoles in a dielectric spheroid of constant refractive index $n=3$, with dimensions $R_x=240\,$nm, $R_y=120\,$nm, and $R_z=120\,$nm, in air $n_{\text{env}}=1$. 
			Plotted are the squared sum of the tensor norms of all generalized polarizability tensors for each multipole order. 
			This is equivalent to the maximal possible radiated energy by the specific multipole moment, assuming it is optimally excited. 
			From left to right: total dipole moment (blue), magnetic dipole (green), total electric quadrupole (orange) and magnetic quadrupole (red).
			Insets show $xy$-slices (left subplots) and $yz$-slices (right subplots) through the spheroid's center of the electric field intensity and the $\mathbf{E}$-field real part (arrows) upon optimum excitation of the respective modes. The wavelengths are indicated by vertical dashed lines in the spectra.
		}
		\label{fig:spheroid_eigenmode_spectrum}
	\end{figure*}
	
	\begin{figure*}[t]
		\centering
		\includegraphics[width=\textwidth]{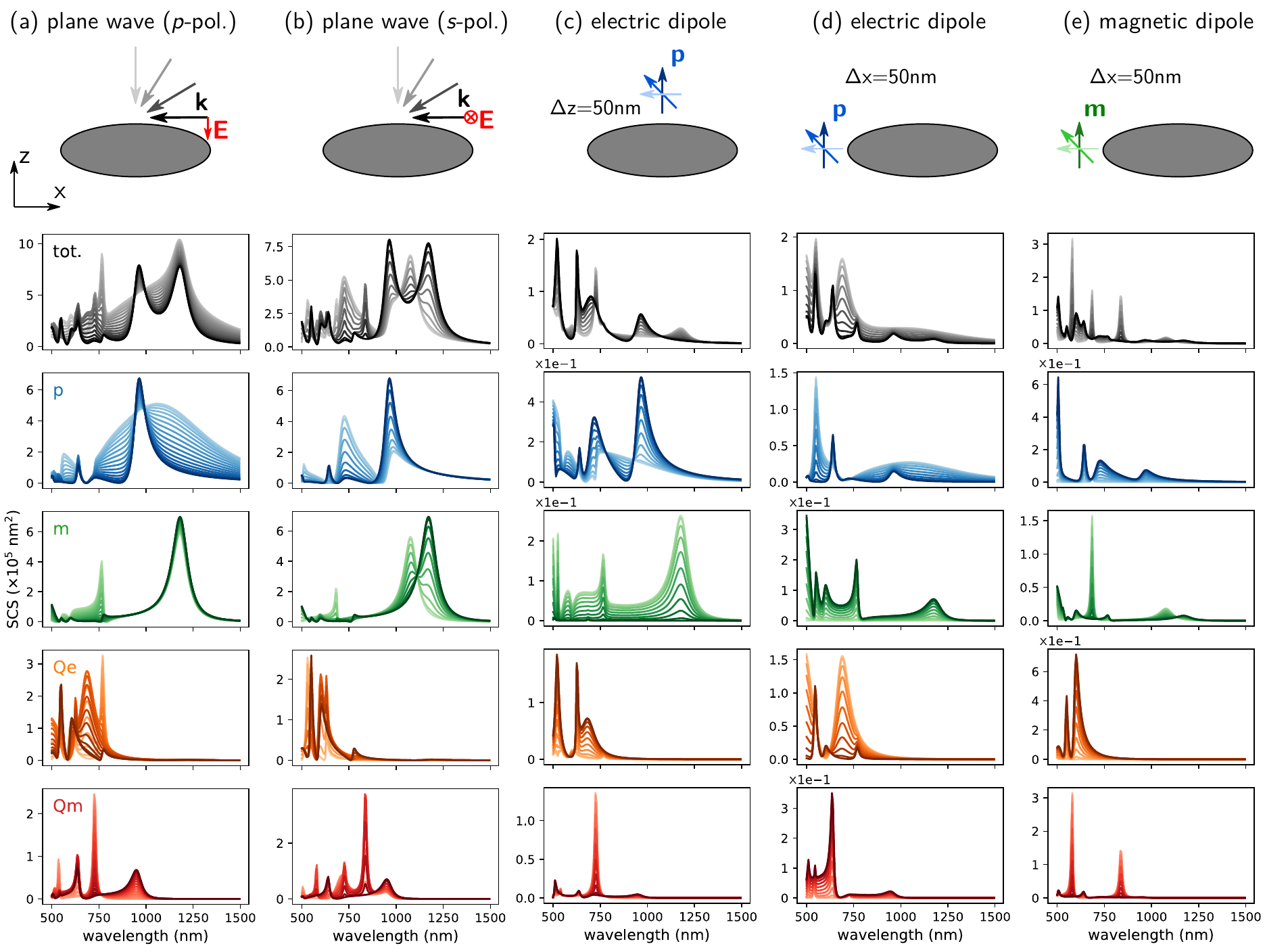}
		\caption{
			Scattering spectra for the same structure as in figure~\ref{fig:spheroid_eigenmode_spectrum}.
			Via the generalized polarizabilities the modal decomposition of the scattering intensity for various different illuminations can be calculated very efficiently.
			We compare here the multipole expansion for various illuminations. (a) $p$-polarized plane wave with incident directions varying from $x$ (dark shades) to $z$ (light shades). (b) same as (a) but for $s$ polarization. 
			(c) illumination by a local electric dipole emitter, $\Delta z=50\,$nm above the spheroid top surface center. Light to dark shades indicate the transition of the dipole orientation from $x$ to $z$.
			(d) same as (c) but the dipole is placed by $\Delta x=50\,$nm at the left outside the spheroid.
			(e) same as (e) but with a magnetic dipole transition as local light source.
			Plane wave illumination spectra show the scattering cross section. Dipole illumination spectra show the scattered intensity in arbitrary units.
		}
		\label{fig:spheroid_variable_illuminations}
	\end{figure*}

	The generalized polarizability tensors reproduce precisely the exact multipole expansion of the electric polarization density for whichever illumination. 
	We demonstrate this with a large dielectric cuboid of lossless material with constant refractive index $n=3.5$ and dimensions $W\times L\times H = 240\times 240\times 330\,$nm$^3$, placed in air ($n_{\text{env}}=1$). 
	The full multipole expansion for illumination with a local source, and a comparison with the long wavelength approximation are given in the supporting information (SI) figures~S1-S3 \checkref{supinfo}).
%
	In figure~\ref{fig:spectra_radpattern} we show scattering under plane wave illumination with oblique incident angle of $135^{\circ}$, and linear $p$-polarization.
	The six polar plots in the top of figure~\ref{fig:spectra_radpattern} show the radiation patterns of the scattered field in the scattering plane for several wavelengths. 
	Solid black lines correspond to the result of full-field calculations, dashed purple lines are obtained via the multipoles from the generalized polarizabilities.\cite{jacksonClassicalElectrodynamics1999, miroshnichenkoNonradiatingAnapoleModes2015}
	We find an excellent agreement, only slight differences can be spotted in particular for the shorter wavelengths where higher order modes begin to contribute.
	Integrating the intensity over the full $4\pi$ solid angle reveals an almost perfect quantitative agreement between full simulations and multipole model, as long as the quadrupolar order of the multipole expansion is sufficient to describe the scattering (here for $\lambda\gtrsim 600$\,nm). 
	In the bottom plot of figure~\ref{fig:spectra_radpattern}, solid lines correspond to the multipole moments calculated from the full simulation, while dashed lines correspond to the generalized polarizability formalism.
	In the SI Fig.~S3,\checkref{supinfo} the same spectra and radiation patterns are shown using the long-wavelength approximation for the generalized polarizabilities.
	
	\section{Density of multipole modes and impact of illumination conditions}
	
	Often, scattering spectra obtained with a fixed incident field (typically a plane wave) are used to characterize the optical properties of nanostructures. As explained previously, especially in dielectric nanostructures a large variety of multipole modes can exist. 
	Their actual appearance, however, is strongly dependent also on the illumination.
	While in particles of high symmetry pure multipole modes can be addressed using cylindrical vector beams,\cite{dasBeamEngineeringSelective2015, montagnacEngineeredFarfieldOptical2022} the response of non-symmetric particles is in general not easily predictable.
	In scattering spectra with fixed illumination, various multipole modes of the structure might even remain invisible.
	The generalized polarizabilities, however, intrinsically contain all possible illuminations. 
	To obtain the entirety of the theoretically available multipole moment at a given wavelength, we can thus sum the Frobenius tensor norms of all meshcells' generalized polarizabilities. 
	The obtained quantity corresponds to the maximally achievable amplitude of the multipole moment, for the case that the local phase distribution is optimally adjusted at any position in the structure. 
	It can thus be regarded as the total density of available multipole modes. 
	The square of this quantity finally, is proportional to the energy radiated by the largest possible multipole moment, hence can be interpreted as the multipole mode's total energy density.

	We demonstrate this in figure~\ref{fig:spheroid_eigenmode_spectrum} for a dielectric spheroid made from a constant refractive index material ($n=3$), with half-axes radii $R_x=240\,$nm and $R_y=R_z=120\,$nm, placed in air ($n_{\text{env}}=1$).
	The spectra show, from left to right, the squared sum of generalized polarizability tensor norms for the total electric dipole (blue line), the magnetic dipole (green line), the total electric quadrupole (orange line) and the magnetic quadrupole (red line). 
	Without a detailed analysis, we can recognize a large number of resonant peaks in the different spectra. 
	Note that we can also access the \textit{partial} multipole densities, corresponding to specific components of the multipole moments (e.g. $p_x$, $m_y$ or $Q_e^{xy}$). For the dipole moments for instance, the partial mode density can be obtained by summation of the norms of the corresponding column vectors of the generalized polarizabilities. The distinct spectra for all isolated multipole tensor components are shown in the SI figure~S4. \checkref{supinfo}
	The insets in figure~\ref{fig:spheroid_eigenmode_spectrum} show the electric field intensity maps on slices through the $xy$ and $yz$ planes, after excitation of the multipole modes at the wavelengths that are indicated by dashed lines. The illumination to excite the respective modes is determined by the generalized polarizability tensors, as described in detail further below.
	
	When comparing the different multipoles, we find that several modes are actually not independent. 
	For instance the electric dipole $p_y$ at $\lambda_0=950\,$nm comes with a magnetic quadrupole moment $Q_m^{xz}$. 
	The magnetic dipole $m_z$ at $\lambda_0=770$\,nm consists actually of two in-phase vortices, that simultaneously induce an electric quadrupole moment $Q_e^{xy}$, etc.
	These correlations between the contributions is a result of the fixed expansion basis. The multipole modes are not an orthogonal basis for the description of the fields in non-spherical nano-structures. 
	For an expansion in an orthogonal basis, quasi-normal modes would need to be extracted, which are unique for every nano-resonator.
	However, using the pre-defined set of multipoles is in several ways more convenient. It allows for instance to draw direct analogies with Mie resonances in spherical resonators, the analysis and re-propagation of the modes is straightforward, and we do not need to worry about normalization.

	In figure~\ref{fig:spheroid_variable_illuminations} we now study the same dielectric spheroid under various illuminations. 
	Figures~\ref{fig:spheroid_variable_illuminations}a and~\ref{fig:spheroid_variable_illuminations}b show scattering spectra for $p$-polarized, respectively $s$-polarized plane wave illumination. 
	Different incident angles are indicated by different shades of the plot colors, the lightest shade corresponds to an incidence along $Z$, the darkest shade to an incidence along $X$.
	Figures~\ref{fig:spheroid_variable_illuminations}c and~\ref{fig:spheroid_variable_illuminations}d show spectra of the scattered intensity upon illumination by an electric dipole placed on top, respectively at the side of the spheroid. Here the different shades of the plots indicate the dipole orientation from along $X$ (light colors) to $Z$ (dark colors).
	Figure~\ref{fig:spheroid_variable_illuminations}e finally shows the multipole expansion for the spheroid illuminated by a magnetic dipole emitter at its side, where the color shades indicate again the emitter orientation from $x$- (light colors) to $z$-direction (dark colors).
	
	A comparison of figures~\ref{fig:spheroid_eigenmode_spectrum} and~\ref{fig:spheroid_variable_illuminations} shows, that the fundamental field distribution plays a crucial role in the excitation of the available modes. 
	It demonstrates that a careful choice of the incident field allows to address specific modes of a nanostructure.
	We see for example that the strong field gradients from a dipolar emitter placed very close to the nano-spheroid, excite more efficiently higher order multipoles, than homogeneous fields like a plane wave.

	\begin{figure*}[t]
		\centering
		\includegraphics[width=\textwidth]{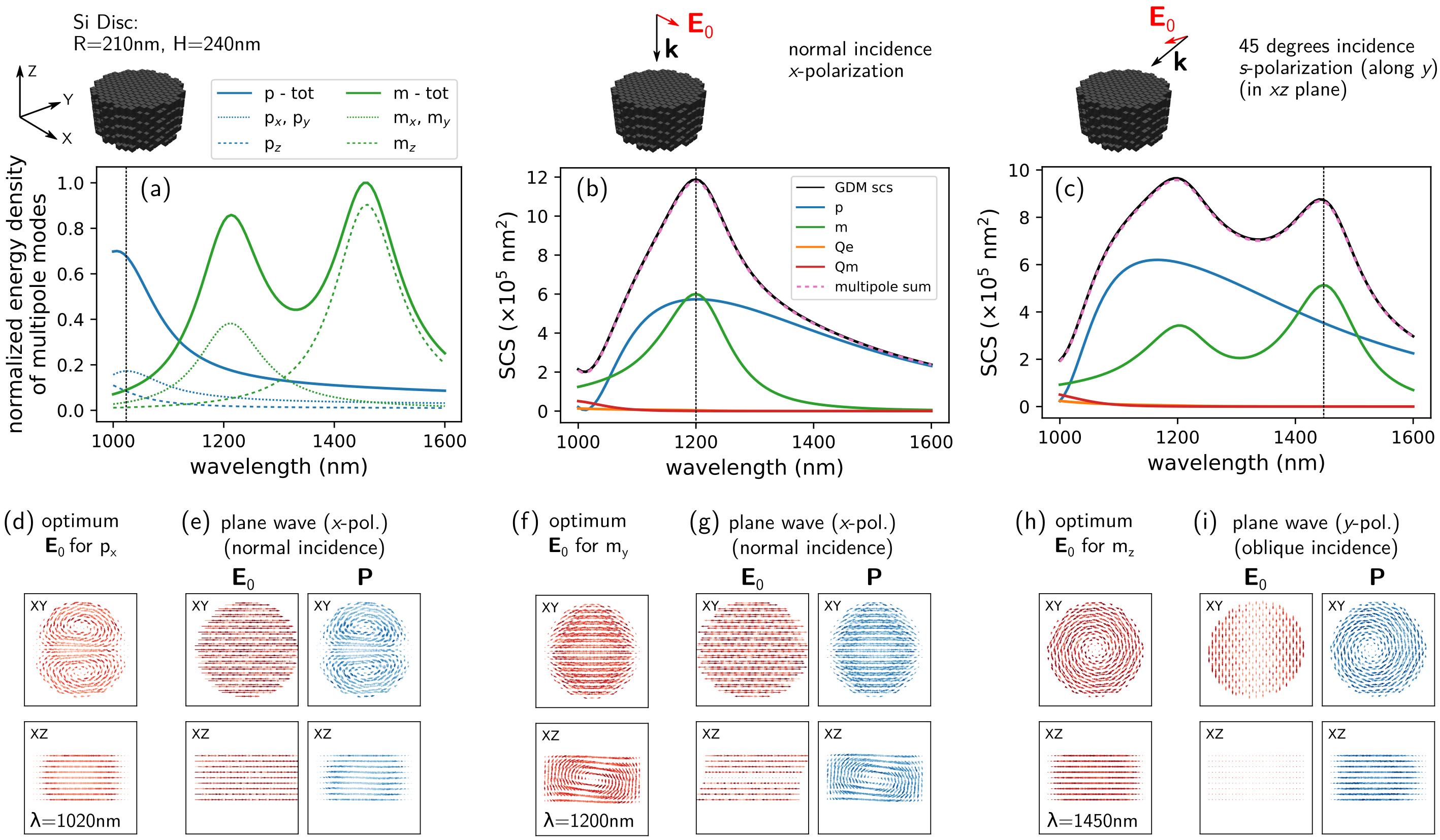}
		\caption{
			Generalized polarizability tensors as visualization tool for the local coupling strength of the illumination to different multipole moments of a silicon disc ($R=210\,$nm, $H=240\,$nm) in air ($n_{\text{env}}=1$). 
			(a) sum of the tensor norms of the generalized polarizabilities of electric (blue) and magnetic (green) dipole moments. The integrated norm of the column vectors corresponding to the Cartesian components of the dipole moments are shown as thin dotted ($x$ and $y$ components) and dashed ($z$ component) lines.
			(b) Scattering spectra and according multipole decomposition for normal illumination with a linearly $x$-polarized plane wave.
			(c) same as in (b) but with oblique incidence ($45^{\circ}$, $s$-polarized)
			(d) Red: Optimum illumination field distributions at the wavelength $1020\,$nm, with maximum possible $x$ and $y$ electric dipole moment.
			(e) Plane wave illumination (red) and according internal field distribution (blue).
			(f-g) like (d-e) but showing the optimum illumination field to induce a magnetic $m_y$ dipole moment at $\lambda_0=1200$\,nm (f) and the internal field induced by a normal incidence plane wave (g).
			(h-i) Ideal field to excite a magnetic $m_z$ moment (h) compared to the internal field induced by an oblique $s$-polarized plane wave (i). 
		}
		\label{fig:local_coupling_strength}
	\end{figure*}

	\section{Analysis of dielectric Huygens sources}

	The generalized polarizabilities of each mesh-cell in a discretized nanostructure correspond to the local strength with which an illumination electric field $\mathbf{E}_0$ induces an according multipole moment in the nanostructure.
	The spatial distribution of the generalized polarizability tensors can therefore be interpreted as the local coupling efficiency of an incident field to the according multipole moments. 
	If one managed to shape an illumination field to correspond exactly to the generalized polarizability distribution, such field would ideally induce the according multipole moment in the structure. 
	In consequence, the generalized polarizability can be used to visualize the spatial zones of strong interaction between an illumination and the multipole moments.

	\paragraph*{Dielectric nanostructures -- Huygens sources}
	We illustrate the possibility of such an analysis by the example of a silicon disc with radius $R=210\,$nm and height $H=240\,$nm, corresponding to the type of structure recently proposed by Decker et al. as unit cell for dielectric Huygens' metasurfaces.\cite{deckerHighEfficiencyDielectricHuygens2015} 
	A Huygens' metasurface exploits the so-called Kerker effect resulting in forward-only scattering, to achieve unitary transmission.\cite{kerkerElectromagneticScatteringMagnetic1983, pfeifferMetamaterialHuygensSurfaces2013, wiechaStronglyDirectionalScattering2017, marcoBroadbandForwardLight2021}
	The permittivity of silicon is taken from literature,\cite{edwardsSiliconSi1997} the disc is placed in air.
	
	The disc dimensions are chosen such that under normal incidence plane wave illumination, the electric and magnetic dipole resonances spectrally overlap and have similar magnitude (see figure~\ref{fig:local_coupling_strength}b).
	In figure~\ref{fig:local_coupling_strength}a we first show the mode densities for the electric and magnetic dipoles. 
	As expected, the disc symmetry leads to degenerate $p_x$ and $p_y$ as well as $m_x$ and $m_y$ modes with resonances at $\lambda_0=1020\,$nm, respectively $\lambda_0=1200\,$nm. 
	Furthermore, we see a strong magnetic dipole mode $m_z$ around $\lambda_0=1450\,$nm.
	Surprisingly, compared to the mode density spectrum, under plane wave illumination the electric dipole has its maximum shifted by around $200\,$nm and coincides with the magnetic dipole. Furthermore, we see that the $m_z$ mode is not contributing to the scattering spectrum under normal incidence.
	However, the $m_z$ mode can be addressed using a plane wave at a $45^{\circ}$ oblique incident angle and $s$-polarization (electric field parallel to the disc top surface), as shown in figure~\ref{fig:local_coupling_strength}c.

	To understand these observations we now have a look at the illumination fields that ideally induce the respective multipoles. These ideal fields are directly obtained from the generalized polarizabilities and we compare them to the internal field induced by a plane wave.
	In figure~\ref{fig:local_coupling_strength}d (red quiver plot) we show the $x$-column vector of the generalized polarizabilities. This represents the illumination which maximally excites the $p_x$ dipole moment at $\lambda_0=1020\,$nm. 
	Figure~\ref{fig:local_coupling_strength}e shows (blue quiver plot) that a normally incident plane wave induces an anapole, known to couple very inefficiently to far-field scattering because of destructive interference of internal field regions with opposite phase.\cite{miroshnichenkoNonradiatingAnapoleModes2015}
	This explains why a maximum in the mode density can occur at a minimum in the scattering spectrum (c.f. blue lines in Figs~\ref{fig:local_coupling_strength}a-b).
	Going back to the optimum field for $p_x$ excitation (Fig.~\ref{fig:local_coupling_strength}d. 
	In contrast to a plane wave this has a phase distribution that matches the anapole and in fact induces a strong electric dipole moment, which efficiently couples to the far-field (see also SI figure~S5). \checkref{supinfo})
	
	By having a look at the column-vectors of the generalized polarizability tensors for the magnetic dipole, we find that the magnetic $m_y$ moment at $\lambda=1200\,$nm can be ideally excited with an illumination field that has a vortex in the $XZ$ plane (see bottom panel in figure~\ref{fig:local_coupling_strength}f). 
	An $x$-polarized normally incident plane wave has field components with opposite phase at top and bottom of the silicon disc (figure~\ref{fig:local_coupling_strength}g). 
	At the upper and lower facets of the Si disc, this is in accord with the ideal field and thus couples well to the magnetic dipole $m_y$. 
	Note that also a side-wards (e.g. along $X$) incident plane wave with polarization along $Z$ would couple to the $m_y$ magnetic dipole component, via the electric field components $E_{0,z}$ of opposite phase at the left an right sides.
	
	In figure~\ref{fig:local_coupling_strength}h we finally show the optimum illumination field for excitation of an $m_z$ dipole moment at the wavelength $\lambda=1450\,$nm, and find that it corresponds to a field-vortex in the $XY$ plane. 
	An $s$-polarized oblique plane wave (polarization along $Y$) has an appropriate phase difference at the left and right side of the silicon disc, and thus induces the same dipole moment (see figure~\ref{fig:local_coupling_strength}i).
	Note, that the vortex-like ideal field distribution is the reason why a magnetic dipole along $z$ can be excited efficiently by an azimuthally polarized, focused vectorbeam.\cite{mannaSelectiveExcitationEnhancement2020, montagnacEngineeredFarfieldOptical2022}
	Scattering spectra of the Si disc illuminated by the optimum excitation fields shown in figures~\ref{fig:local_coupling_strength}d,~\ref{fig:local_coupling_strength}f and~\ref{fig:local_coupling_strength}h are shown in the SI figure~S5. \checkref{supinfo}

	\paragraph*{Impact on Huygens metasurfaces}
	
	The strong dependence on the illumination of the dipole modes in dielectric nanostructures has important implications for their usage as elementary blocks in Huygens metasurfaces, as Gigli et al. have already recently discussed.\cite{gigliFundamentalLimitationsHuygens2021} 
	During the design procedure of a metasurface, a lookup table is created, for which the phase-delays of various meta-atoms are simulated. 
	These simulations are usually done with periodic boundary conditions and using a fixed illumination angle. The phase delays of the meta-atoms are subsequently matched with the target metasurface phase map and the structures are placed accordingly.
	The resulting metasurface obviously does not have the periodicity, that was assumed for the simulations. 
	In consequence the local fields are perturbed by the non-periodic structure arrangement. 
	
	The crucial point is now, that a variation of the local illumination can easily lead to the unexpected excitation of a mode that may be ``invisible'' for a plane wave, as for instance the $m_y$ and $m_z$ dipole moments in our above analysis.
	Furthermore, as we found in the precedent section, the broad electric dipole resonance under normal plane wave illumination (Fig.~\ref{fig:local_coupling_strength}b) is in fact no eigenmode of the system, but rather a dressed mode, dressed by the plane wave illumination. A local source will interact very differently with the structure (c.f. Fig.~\ref{fig:spheroid_variable_illuminations}).
	Also a rotation of the electric or magnetic dipole moment's orientation can naturally occur if the effective incident angle locally deviates from the plane wave, due to scattering from surrounding structures. 
	Also the relative magnitude between the electric and magnetic dipole moments can be significantly affected, as can be seen for instance around $\lambda=1200\,$nm, when comparing figures~\ref{fig:local_coupling_strength}b and~\ref{fig:local_coupling_strength}c.
	In consequence Kerker's condition will not be satisfied anymore. Reflection will occur, reducing the efficiency of the Huygens metasurface. 
	In conclusion, a Huygens metasurface based on dielectric nanoresonators requires very delicate optimization of each single constituent, to match the local environment.

	\section{Conclusions}
	
	In summary, by combining the exact multipole decomposition with the concept of a generalized propagator, we derived expressions for what we call \textit{generalized polarizabilities}. 
	These are defined for each meshcell of a volume discretized nanostructure and describe the contribution of the respective meshcell to the induced multipole moment.
	The generalized polarizability tensors allow to obtain at basically no computational cost the exact multipole expansion of the optical response of a nanostructure for arbitrary illuminations and they allow to calculate spectra of the total density of multipole modes, independent of a specific illumination. 
	The formalism can also be used as a tool for direct visualization of the local coupling strength of an illumination field to the different multipole moments. 
	This is interesting for instance for beam-shaping experiments where the nature of the induced optical response in a nanostructure may be controlled through a complex illumination field.\cite{volpeControllingOpticalField2009, wozniakSelectiveSwitchingIndividual2015, dasBeamEngineeringSelective2015}
	We foresee in particular relevant applications in electron microscopy.\cite{guzzinatiProbingSymmetryPotential2017, alexanderNearFieldMappingPhotonic2021}
	We believe that the mode-density analysis via our generalized polarizabilities formalism will be a very valuable tool, for example to anticipate the robustness of a dielectric nanostructure as a meta-atom in a Huygens metasurface.
	Finally, we anticipate that the very low storage requirements will allow to use the generalized polarizabilities efficiently in lookup tables and also together with deep learning for various applications ranging from the interpretation of the optical properties of individual nanostructures to the design of complex metasurfaces.\cite{wiechaDeepLearningMeets2020, anDeepConvolutionalNeural2021, wiechaDeepLearningNanophotonics2021, majorelDeepLearningEnabled2022}

	\begin{acknowledgments}
		We thank Aurélien Cuche and Otto L. Muskens for fruitful discussions. 
		A.P. acknowledges support by Airbus Defence and Space (ADS), through a Ph.D. CIFRE fellowship (No. 2008/0925).
		A.E.-R. thanks the Institute of Quantum Technology in Occitanie IQO and the Université Paul Sabatier Toulouse for an UPS excellence PhD grant.
		This work was supported by the Toulouse HPC CALMIP (grant p20010).
	\end{acknowledgments}

	\section*{Disclosures}
		The authors declare no conflicts of interest.

	\section*{Supporting Informations}
	\begin{itemize}
		\item A pdf providing a comparison of the exact multipole generalized polarizabilities and the long wavelength limit multipole expansion as well as further details on the mode-analysis of dielectric nanostructures.
		\item Example scripts written in python, demonstrating the use of our method, which we implemented in the publicly available open source package \href{https://wiechapeter.gitlab.io/pyGDM2-doc/index.html}{\textit{pyGDM}}.
	\end{itemize}

	\section*{Appendix}
	\subsection{Generalized Field Propagator}\label{appendix:generalized_propagator}

	For an environment which contains some nano-scatterer(s) of electric susceptibility $\overline{\overline{\boldsymbol{\chi}}}(\mathbf{r}, \omega)$, occupying the volume $V_{ns}$, let us define the generalized field propagator $\overline{\overline{\mathbf{K}}}(\mathbf{r}, \mathbf{r}', \omega)$ as the tensor that links the illumination electric field $\mathbf{E}_0(\mathbf{r}', \omega)$ at $\mathbf{r}'$ with the total field $\mathbf{E}(\mathbf{r}, \omega)$ at $\mathbf{r}$:\cite{martinGeneralizedFieldPropagator1995}
	\begin{equation}
		\mathbf{E}(\mathbf{r}, \omega) = \overline{\overline{\mathbf{K}}}(\mathbf{r}, \mathbf{r}', \omega) \cdot \mathbf{E}_0(\mathbf{r}', \omega)\, .
	\end{equation}
	$\overline{\overline{\mathbf{K}}}$ includes scattering as well as possible absorption by the nano-scatterer(s).

	For nanostructures of arbitrary shape and material distribution, it is in general not possible to solve the scattering problem analytically and a numerical approach is required. 
	We therefore start by discretizing the volume integral over the nanostructure in the Lippmann-Schwinger equation, Eq.~\eqref{eq:lippmann_schwinger}.
	To do this, we subdivide the volume of the structure into $N$ unit cells, located at positions $r_i$ on a regular grid. The differential term $\drpp$ is replaced by the unit cell volume $V_{c}$. 
	Technically the procedure is identical with the transition from equation~\eqref{eq:generalized_propa} to equation~\eqref{eq:generalized_propa_discretized} in the main text, and we obtain:
	\begin{multline}\label{eq:lippmann_schwinger_discretized}
		\mathbf{E}(\mathbf{r}_i, \omega) = \mathbf{E}_0(\mathbf{r}_i, \omega) + \\
		+ \sum_{j=1}^N\ \overline{\overline{\mathbf{G}}}(\mathbf{r}_i, \mathbf{r}_j, \omega) \cdot \overline{\overline{\boldsymbol{\chi}}}(\mathbf{r}_j, \omega) \cdot \mathbf{E}(\mathbf{r}_j, \omega)V_{c}\, .
	\end{multline}
	By defining two super-vectors of length $3N$, $\mathbf{E}_{0,\text{obj}}$ and $\mathbf{E}_{\text{obj}}$, containing the electromagnetic fields at each unit cell's position $\mathbf{r}_j$, the expression~\eqref{eq:lippmann_schwinger_discretized} can be written in matrix form:
	\begin{equation}\label{eq:matrix_form_coupling_LS}
		\mathbf{E}_{0,\text{obj}} = \overline{\overline{\mathbf{M}}}\cdot\mathbf{E}_{\text{obj}} \, .
	\end{equation}
	The $(3N\times3N)$ matrix $\overline{\overline{\mathbf{M}}}$ is composed of $(3\times3)$ matrices, depicting the pairwise coupling between all $N$ unit cells.
	From comparison of equations~\eqref{eq:lippmann_schwinger_discretized} and~\eqref{eq:matrix_form_coupling_LS}, we obtain the following form for these matrices:
	\begin{equation}
		\mathbf{M} (\mathbf{r}_i, \mathbf{r}_j) = \mathbf{I} - \sum_{j=1}^N\ \overline{\overline{\mathbf{G}}}(\mathbf{r}_i, \mathbf{r}_j, \omega) \cdot \overline{\overline{\boldsymbol{\chi}}}(\mathbf{r}_j, \omega) \cdot \mathbf{E}(\mathbf{r}_j, \omega)V_{c}\, ,
	\end{equation}
	where $\mathbf{I}$ is the unit tensor.
	By inverting the matrix $\overline{\overline{\mathbf{M}}}$, we obtain the (discretized) generalized field propagators for positions inside the nanostructure: 
	\begin{equation}
		\overline{\overline{\mathbf{K}}}(\mathbf{r}_i, \mathbf{r}_j, \omega) = \overline{\overline{\mathbf{M}}}^{-1}_{i,j} \, ,
	\end{equation}
	where the indices $i$ and $j$ indicate the $(i,j)$-ieth $(3,3)$ submatrix of the inverted matrix~$\overline{\overline{\mathbf{M}}}^{-1}$.

	\subsection{Quadrupole generalized polarizabilities}\label{appendix:quadrupole_terms}
	The $ab$-component of the exact electric and magnetic quadrupole moments writes:\cite{alaeeElectromagneticMultipoleExpansion2018}
	
	\begin{subequations}\label{eq:quadrupoles_p}
		\begin{align}
		Q_{e}^{ab}(\mathbf{r}_0, \omega) &= Q_{e0}^{ab}(\mathbf{r}_0, \omega) + Q_{et}^{ab}(\mathbf{r}_0, \omega)
		\\
		Q_{e0}^{ab}(\mathbf{r}_0, \omega) &= 3 \int\limits_{V_{ns}} \drp
		\bigg[ 3\Big( r'^{b} P^a + r'^{a}P^b \Big) \\
		&\quad\quad - 2\, \Big( \mathbf{r}' \cdot \Prp \Big) \delta^{a b} \bigg]
		\frac{j_1 (kr')}{kr'}
		\nonumber
		\\
		Q_{et}^{ab}(\mathbf{r}_0, \omega) &= 6 k^2 V_{c}^2 \int\limits_{V_{ns}} \drp
		\Bigg[ 5 r'^{a}r'^{b} \Big( \mathbf{r}' \cdot \Prp \Big) 
		\\
		& \quad\quad - \Big( r'^{b} P^a + r'^{a}P^b \Big) r_i^2 
		\nonumber
		\\ & \quad\quad 
		- r_i^2\, \Big( \mathbf{r}' \cdot \Prp \Big) \delta^{ab} \Bigg]
		\frac{j_3 (kr')}{(kr')^3}
		\nonumber
		\end{align}
	\end{subequations}
	\begin{equation}\label{eq:quadrupoles_m}
	\begin{aligned}
	Q_{m}^{ab}(\mathbf{r}_0, \omega) &= -15 \text{i}k \int\limits_{V_{ns}} \drp
	\bigg[ r'^a \Big( \mathbf{r}' \times \Prp \Big)^b \\
	&\quad\quad + r'^b \Big( \mathbf{r}' \times \Prp \Big)^a \bigg]
	\frac{j_2 (kr')}{(kr')^2}
	\end{aligned}
	\end{equation}
	$\delta^{ab}$ is the Kronecker symbol and $P^a$ is the $a$-component of the vector $\Prp$.
	$\overline{\overline{Q}}_{e0}$ and $\overline{\overline{Q}}_{et}$ are, respectively, the first order term, and the toroidal quadrupole term of the total electric quadrupole moment $\overline{\overline{Q}}_{e}$.
	
	After discretization, substitution with Eq.~\eqref{eq:generalized_propa_discretized}, and re-ordering of the summations, we find for the electric quadrupole terms:
	\begin{subequations}\label{eq:electric_quadrupole_gen_propagator_interchanged_summation}
		\begin{align}
		Q_{e0}^{ab}(\mathbf{r}_0, \omega) &= 3 V_{c}^2 \sum\limits_{l=1}^3 \sum\limits_{j}^{N_c}\ 
		\Bigg[ \sum\limits_{i}^{N_c} 
		\Bigg( 3\frac{1}{3} \Big( r_{i}^{b} \chi_{i,\gamma}^{a} K_{ij,\epsilon}^{\gamma} + r_{i}^{a} \chi_{i,\gamma}^{b} K_{ij,\epsilon}^{\gamma} \Big) 
		\nonumber \\
		& \quad\quad - 2\, \delta^{a b} \Big( r_{i,l} \chi_{i,\gamma}^{l} K_{ij,\epsilon}^{\gamma} \Big) \Bigg)
		\frac{j_1 (kr_i)}{kr_i}
		\Bigg] E_{0j}^{\epsilon}
		\\
		Q_{et}^{ab}(\mathbf{r}_0, \omega) &= 6 k^2 V_{c}^2 \sum\limits_{l=1}^3 \sum\limits_{j}^{N_c}\ 
		\Bigg[ \sum\limits_{i}^{N_c} 
		\Bigg( 5 r_i^{a}r_i^{b} \Big( r_{i,l} \chi_{i,\gamma}^{l} K_{ij,\epsilon}^{\gamma} \Big) 
		\nonumber \\
		& \quad\quad - \frac{r_i^2}{3}\, \Big( r_{i}^{b} \chi_{i,\gamma}^{a} K_{ij,\epsilon}^{\gamma} + r_{i}^{a} \chi_{i,\gamma}^{b} K_{ij,\epsilon}^{\gamma} \Big) 
		\\
		& \quad\quad - r_i^2\, \delta^{a b} \Big( r_{i,l} \chi_{i,\gamma}^{l} K_{ij,\epsilon}^{\gamma} \Big) \Bigg)
		\frac{j_3 (kr_i)}{(kr_i)^3}
		\Bigg] E_{0j}^{\epsilon}\nonumber
		\end{align}
	\end{subequations}

	\noindent
	The sum over the index $l$ is again introduced to be able to perform the scalar product $\mathbf{r}_i\cdot \mathbf{P}_i$ after summation over the index $i$.
	The terms in square brackets in Eqs.~\eqref{eq:electric_quadrupole_gen_propagator_interchanged_summation} correspond to the according $N_c$ electric quadrupolar generalized polarizabilities:
	
	\begin{subequations}\label{eq:electric_quadrupole_gen_polarizabilities}
		\begin{align}
		\alpha^{Q_{e0}, ab}_{j,l\epsilon}(\mathbf{r}_0, \omega) &= 3 V_{c}^2
		\sum\limits_{i}^{N_c} 
		\Bigg( 3\frac{1}{3} \Big( r_{i}^{b} \chi_{i,\gamma}^{a} K_{ij,\epsilon}^{\gamma} + r_{i}^{a} \chi_{i,\gamma}^{b} K_{ij,\epsilon}^{\gamma} \Big) 
		\nonumber 
		\\
		& \quad\quad - 2\, \delta^{a b} \Big( r_{i,l} \chi_{i,\gamma}^{l} K_{ij,\epsilon}^{\gamma} \Big) \Bigg)
		\frac{j_1 (kr_i)}{kr_i} 
		\\
		\alpha^{Q_{et}, ab}_{j,l\epsilon}(\mathbf{r}_0, \omega) &= 6 k^2 V_{c}^2 
		\sum\limits_{i}^{N_c} 
		\Bigg( 5 r_i^{a}r_i^{b} \Big( r_{i,l} \chi_{i,\gamma}^{l} K_{ij,\epsilon}^{\gamma} \Big) 
		\\
		& \quad\quad - \frac{r_i^2}{3}\, \Big( r_{i}^{b} \chi_{i,\gamma}^{a} K_{ij,\epsilon}^{\gamma} + r_{i}^{a} \chi_{i,\gamma}^{b} K_{ij,\epsilon}^{\gamma} \Big) 
		\nonumber 
		\\
		& \quad\quad - r_i^2\, \delta^{a b} \Big( r_{i,l} \chi_{i,\gamma}^{l} K_{ij,\epsilon}^{\gamma} \Big) \Bigg)
		\frac{j_3 (kr_i)}{(kr_i)^3} \nonumber
		\end{align}
	\end{subequations}
	which can be used to calculate the electric quadrupole for any illumination $\mathbf{E}_0$ as
	%
	\begin{equation}
		\overline{\overline{Q}}_{e}(\mathbf{r}_0, \omega) =
		\overbrace{\sum\limits_{l=1}^3 \sum\limits_{j}^{N_c} \overline{\overline{\alpha}}^{Q_{e0}}_{j,l} \cdot \mathbf{E}_{0j}}^{\text{quadrupole}}\ +\ 
		\overbrace{\sum\limits_{l=1}^3 \sum\limits_{j}^{N_c} \overline{\overline{\alpha}}^{Q_{et}}_{j,l} \cdot \mathbf{E}_{0j}}^{\text{toroidal quadrupole}}
	\end{equation}

	\noindent
	Analogously, the magnetic quadrupole can be written as:
	
	\begin{multline}\label{eq:magnetic_quadrupole_gen_propagator_interchanged_summation}
		Q_{m}^{ab}(\mathbf{r}_0, \omega) = -15\text{i}k V_c^2 \sum\limits_{j}^{N_c} \Bigg[ \sum\limits_{i}^{N_c} 
		\Big( r_i^a \epsilon^b_{\zeta\kappa}\ r_i^{\zeta} \chi_{i,\gamma}^{\kappa} K_{ij,\epsilon}^{\gamma} + \\
		r_i^b \epsilon^a_{\zeta\kappa}\ r_i^{\zeta} \chi_{i,\gamma}^{\kappa} K_{ij,\epsilon}^{\gamma}\Big) 
		\frac{j_2(k r_i)}{(kr_i)^2}\Bigg] E_{0,j}^{\epsilon}
	\end{multline}
	\noindent
	leading to the following definition of the magnetic quadrupole generalized polarizabilities:
	
	\begin{multline}\label{eq:magnetic_quadrupole_gen_polarizabilities}
		\alpha^{Q_m, ab}_{j,\epsilon}(\mathbf{r}_0, \omega) = -15\text{i}k V_c^2 
		\sum\limits_{i}^{N_c} 
		\Big( r_i^a \epsilon^b_{\zeta\kappa}\ r_i^{\zeta} \chi_{i,\gamma}^{\kappa} K_{ij,\epsilon}^{\gamma} + \\
		r_i^b \epsilon^a_{\zeta\kappa}\ r_i^{\zeta} \chi_{i,\gamma}^{\kappa} K_{ij,\epsilon}^{\gamma}\Big) 
		\frac{j_2(k r_i)}{(kr_i)^2}
	\end{multline}
	\noindent
	from which we can now calculate the magnetic quadrupole moment for any illumination field:
	
	\begin{equation}
		\overline{\overline{Q}}_{m}(\mathbf{r}_0, \omega) = \sum\limits_{j}^{N_c} \overline{\overline{\alpha}}^{Q_m}_{j} \cdot \mathbf{E}_{0j}
	\end{equation}
	
	Note that due to the scalar products occurring in Eqs.~\eqref{eq:electric_quadrupole_gen_propagator_interchanged_summation}, we obtain $N_c$ rank 4 tensors as electric quadrupole generalized polarizabilities (additional index $l$). The magnetic quadrupole on the other hand can be expressed by $N_c$ rank 3 tensors.


\begin{thebibliography}{76}%
\makeatletter
\providecommand \@ifxundefined [1]{%
 \@ifx{#1\undefined}
}%
\providecommand \@ifnum [1]{%
 \ifnum #1\expandafter \@firstoftwo
 \else \expandafter \@secondoftwo
 \fi
}%
\providecommand \@ifx [1]{%
 \ifx #1\expandafter \@firstoftwo
 \else \expandafter \@secondoftwo
 \fi
}%
\providecommand \natexlab [1]{#1}%
\providecommand \enquote  [1]{``#1''}%
\providecommand \bibnamefont  [1]{#1}%
\providecommand \bibfnamefont [1]{#1}%
\providecommand \citenamefont [1]{#1}%
\providecommand \href@noop [0]{\@secondoftwo}%
\providecommand \href [0]{\begingroup \@sanitize@url \@href}%
\providecommand \@href[1]{\@@startlink{#1}\@@href}%
\providecommand \@@href[1]{\endgroup#1\@@endlink}%
\providecommand \@sanitize@url [0]{\catcode `\\12\catcode `\$12\catcode
  `\&12\catcode `\#12\catcode `\^12\catcode `\_12\catcode `\%12\relax}%
\providecommand \@@startlink[1]{}%
\providecommand \@@endlink[0]{}%
\providecommand \url  [0]{\begingroup\@sanitize@url \@url }%
\providecommand \@url [1]{\endgroup\@href {#1}{\urlprefix }}%
\providecommand \urlprefix  [0]{URL }%
\providecommand \Eprint [0]{\href }%
\providecommand \doibase [0]{http://dx.doi.org/}%
\providecommand \selectlanguage [0]{\@gobble}%
\providecommand \bibinfo  [0]{\@secondoftwo}%
\providecommand \bibfield  [0]{\@secondoftwo}%
\providecommand \translation [1]{[#1]}%
\providecommand \BibitemOpen [0]{}%
\providecommand \bibitemStop [0]{}%
\providecommand \bibitemNoStop [0]{.\EOS\space}%
\providecommand \EOS [0]{\spacefactor3000\relax}%
\providecommand \BibitemShut  [1]{\csname bibitem#1\endcsname}%
\let\auto@bib@innerbib\@empty
\bibitem [{\citenamefont {Kuznetsov}\ \emph {et~al.}(2016)\citenamefont
  {Kuznetsov}, \citenamefont {Miroshnichenko}, \citenamefont {Brongersma},
  \citenamefont {Kivshar},\ and\ \citenamefont
  {Luk'yanchuk}}]{kuznetsovOpticallyResonantDielectric2016}%
  \BibitemOpen
  \bibfield  {author} {\bibinfo {author} {\bibfnamefont {A.~I.}\ \bibnamefont
  {Kuznetsov}}, \bibinfo {author} {\bibfnamefont {A.~E.}\ \bibnamefont
  {Miroshnichenko}}, \bibinfo {author} {\bibfnamefont {M.~L.}\ \bibnamefont
  {Brongersma}}, \bibinfo {author} {\bibfnamefont {Y.~S.}\ \bibnamefont
  {Kivshar}}, \ and\ \bibinfo {author} {\bibfnamefont {B.}~\bibnamefont
  {Luk'yanchuk}},\ }\href {\doibase 10.1126/science.aag2472} {\bibfield
  {journal} {\bibinfo  {journal} {Science}\ }\textbf {\bibinfo {volume} {354}}
  (\bibinfo {year} {2016}),\ 10.1126/science.aag2472}\BibitemShut {NoStop}%
\bibitem [{\citenamefont {Fu}\ \emph {et~al.}(2013)\citenamefont {Fu},
  \citenamefont {Kuznetsov}, \citenamefont {Miroshnichenko}, \citenamefont
  {Yu},\ and\ \citenamefont {Luk'yanchuk}}]{fuDirectionalVisibleLight2013}%
  \BibitemOpen
  \bibfield  {author} {\bibinfo {author} {\bibfnamefont {Y.~H.}\ \bibnamefont
  {Fu}}, \bibinfo {author} {\bibfnamefont {A.~I.}\ \bibnamefont {Kuznetsov}},
  \bibinfo {author} {\bibfnamefont {A.~E.}\ \bibnamefont {Miroshnichenko}},
  \bibinfo {author} {\bibfnamefont {Y.~F.}\ \bibnamefont {Yu}}, \ and\ \bibinfo
  {author} {\bibfnamefont {B.}~\bibnamefont {Luk'yanchuk}},\ }\href {\doibase
  10.1038/ncomms2538} {\bibfield  {journal} {\bibinfo  {journal} {Nature
  Communications}\ }\textbf {\bibinfo {volume} {4}},\ \bibinfo {pages} {1527}
  (\bibinfo {year} {2013})}\BibitemShut {NoStop}%
\bibitem [{\citenamefont {Wiecha}\ \emph {et~al.}(2017)\citenamefont {Wiecha},
  \citenamefont {Cuche}, \citenamefont {Arbouet}, \citenamefont {Girard},
  \citenamefont {{Colas des Francs}}, \citenamefont {Lecestre}, \citenamefont
  {Larrieu}, \citenamefont {Fournel}, \citenamefont {Larrey}, \citenamefont
  {Baron},\ and\ \citenamefont
  {Paillard}}]{wiechaStronglyDirectionalScattering2017}%
  \BibitemOpen
  \bibfield  {author} {\bibinfo {author} {\bibfnamefont {P.~R.}\ \bibnamefont
  {Wiecha}}, \bibinfo {author} {\bibfnamefont {A.}~\bibnamefont {Cuche}},
  \bibinfo {author} {\bibfnamefont {A.}~\bibnamefont {Arbouet}}, \bibinfo
  {author} {\bibfnamefont {C.}~\bibnamefont {Girard}}, \bibinfo {author}
  {\bibfnamefont {G.}~\bibnamefont {{Colas des Francs}}}, \bibinfo {author}
  {\bibfnamefont {A.}~\bibnamefont {Lecestre}}, \bibinfo {author}
  {\bibfnamefont {G.}~\bibnamefont {Larrieu}}, \bibinfo {author} {\bibfnamefont
  {F.}~\bibnamefont {Fournel}}, \bibinfo {author} {\bibfnamefont
  {V.}~\bibnamefont {Larrey}}, \bibinfo {author} {\bibfnamefont
  {T.}~\bibnamefont {Baron}}, \ and\ \bibinfo {author} {\bibfnamefont
  {V.}~\bibnamefont {Paillard}},\ }\href {\doibase
  10.1021/acsphotonics.7b00423} {\bibfield  {journal} {\bibinfo  {journal} {ACS
  Photonics}\ }\textbf {\bibinfo {volume} {4}},\ \bibinfo {pages} {2036}
  (\bibinfo {year} {2017})}\BibitemShut {NoStop}%
\bibitem [{\citenamefont {Kats}\ \emph {et~al.}(2012)\citenamefont {Kats},
  \citenamefont {Genevet}, \citenamefont {Aoust}, \citenamefont {Yu},
  \citenamefont {Blanchard}, \citenamefont {Aieta}, \citenamefont {Gaburro},\
  and\ \citenamefont {Capasso}}]{katsGiantBirefringenceOptical2012}%
  \BibitemOpen
  \bibfield  {author} {\bibinfo {author} {\bibfnamefont {M.~A.}\ \bibnamefont
  {Kats}}, \bibinfo {author} {\bibfnamefont {P.}~\bibnamefont {Genevet}},
  \bibinfo {author} {\bibfnamefont {G.}~\bibnamefont {Aoust}}, \bibinfo
  {author} {\bibfnamefont {N.}~\bibnamefont {Yu}}, \bibinfo {author}
  {\bibfnamefont {R.}~\bibnamefont {Blanchard}}, \bibinfo {author}
  {\bibfnamefont {F.}~\bibnamefont {Aieta}}, \bibinfo {author} {\bibfnamefont
  {Z.}~\bibnamefont {Gaburro}}, \ and\ \bibinfo {author} {\bibfnamefont
  {F.}~\bibnamefont {Capasso}},\ }\href {\doibase 10.1073/pnas.1210686109}
  {\bibfield  {journal} {\bibinfo  {journal} {Proceedings of the National
  Academy of Sciences}\ }\textbf {\bibinfo {volume} {109}},\ \bibinfo {pages}
  {12364} (\bibinfo {year} {2012})}\BibitemShut {NoStop}%
\bibitem [{\citenamefont {Rodrigo}\ \emph {et~al.}(2013)\citenamefont
  {Rodrigo}, \citenamefont {Harutyunyan},\ and\ \citenamefont
  {Novotny}}]{rodrigoCoherentControlLight2013}%
  \BibitemOpen
  \bibfield  {author} {\bibinfo {author} {\bibfnamefont {S.~G.}\ \bibnamefont
  {Rodrigo}}, \bibinfo {author} {\bibfnamefont {H.}~\bibnamefont
  {Harutyunyan}}, \ and\ \bibinfo {author} {\bibfnamefont {L.}~\bibnamefont
  {Novotny}},\ }\href {\doibase 10.1103/PhysRevLett.110.177405} {\bibfield
  {journal} {\bibinfo  {journal} {Physical Review Letters}\ }\textbf {\bibinfo
  {volume} {110}},\ \bibinfo {pages} {177405} (\bibinfo {year}
  {2013})}\BibitemShut {NoStop}%
\bibitem [{\citenamefont {Shcherbakov}\ \emph {et~al.}(2014)\citenamefont
  {Shcherbakov}, \citenamefont {Neshev}, \citenamefont {Hopkins}, \citenamefont
  {Shorokhov}, \citenamefont {Staude}, \citenamefont {{Melik-Gaykazyan}},
  \citenamefont {Decker}, \citenamefont {Ezhov}, \citenamefont
  {Miroshnichenko}, \citenamefont {Brener}, \citenamefont {Fedyanin},\ and\
  \citenamefont {Kivshar}}]{shcherbakovEnhancedThirdHarmonicGeneration2014}%
  \BibitemOpen
  \bibfield  {author} {\bibinfo {author} {\bibfnamefont {M.~R.}\ \bibnamefont
  {Shcherbakov}}, \bibinfo {author} {\bibfnamefont {D.~N.}\ \bibnamefont
  {Neshev}}, \bibinfo {author} {\bibfnamefont {B.}~\bibnamefont {Hopkins}},
  \bibinfo {author} {\bibfnamefont {A.~S.}\ \bibnamefont {Shorokhov}}, \bibinfo
  {author} {\bibfnamefont {I.}~\bibnamefont {Staude}}, \bibinfo {author}
  {\bibfnamefont {E.~V.}\ \bibnamefont {{Melik-Gaykazyan}}}, \bibinfo {author}
  {\bibfnamefont {M.}~\bibnamefont {Decker}}, \bibinfo {author} {\bibfnamefont
  {A.~A.}\ \bibnamefont {Ezhov}}, \bibinfo {author} {\bibfnamefont {A.~E.}\
  \bibnamefont {Miroshnichenko}}, \bibinfo {author} {\bibfnamefont
  {I.}~\bibnamefont {Brener}}, \bibinfo {author} {\bibfnamefont {A.~A.}\
  \bibnamefont {Fedyanin}}, \ and\ \bibinfo {author} {\bibfnamefont {Y.~S.}\
  \bibnamefont {Kivshar}},\ }\href {\doibase 10.1021/nl503029j} {\bibfield
  {journal} {\bibinfo  {journal} {Nano Letters}\ }\textbf {\bibinfo {volume}
  {14}},\ \bibinfo {pages} {6488} (\bibinfo {year} {2014})}\BibitemShut
  {NoStop}%
\bibitem [{\citenamefont {Wiecha}\ \emph {et~al.}(2015)\citenamefont {Wiecha},
  \citenamefont {Arbouet}, \citenamefont {Kallel}, \citenamefont {Periwal},
  \citenamefont {Baron},\ and\ \citenamefont
  {Paillard}}]{wiechaEnhancedNonlinearOptical2015}%
  \BibitemOpen
  \bibfield  {author} {\bibinfo {author} {\bibfnamefont {P.~R.}\ \bibnamefont
  {Wiecha}}, \bibinfo {author} {\bibfnamefont {A.}~\bibnamefont {Arbouet}},
  \bibinfo {author} {\bibfnamefont {H.}~\bibnamefont {Kallel}}, \bibinfo
  {author} {\bibfnamefont {P.}~\bibnamefont {Periwal}}, \bibinfo {author}
  {\bibfnamefont {T.}~\bibnamefont {Baron}}, \ and\ \bibinfo {author}
  {\bibfnamefont {V.}~\bibnamefont {Paillard}},\ }\href {\doibase
  10.1103/PhysRevB.91.121416} {\bibfield  {journal} {\bibinfo  {journal}
  {Physical Review B}\ }\textbf {\bibinfo {volume} {91}},\ \bibinfo {pages}
  {121416} (\bibinfo {year} {2015})}\BibitemShut {NoStop}%
\bibitem [{\citenamefont {Girard}\ \emph {et~al.}(1994)\citenamefont {Girard},
  \citenamefont {Dereux},\ and\ \citenamefont
  {Martin}}]{girardTheoreticalAnalysisLightinductive1994}%
  \BibitemOpen
  \bibfield  {author} {\bibinfo {author} {\bibfnamefont {C.}~\bibnamefont
  {Girard}}, \bibinfo {author} {\bibfnamefont {A.}~\bibnamefont {Dereux}}, \
  and\ \bibinfo {author} {\bibfnamefont {O.~J.~F.}\ \bibnamefont {Martin}},\
  }\href {\doibase 10.1103/PhysRevB.49.13872} {\bibfield  {journal} {\bibinfo
  {journal} {Physical Review B}\ }\textbf {\bibinfo {volume} {49}},\ \bibinfo
  {pages} {13872} (\bibinfo {year} {1994})}\BibitemShut {NoStop}%
\bibitem [{\citenamefont {Chaumet}\ and\ \citenamefont
  {{Nieto-Vesperinas}}(2000)}]{chaumetCoupledDipoleMethod2000}%
  \BibitemOpen
  \bibfield  {author} {\bibinfo {author} {\bibfnamefont {P.~C.}\ \bibnamefont
  {Chaumet}}\ and\ \bibinfo {author} {\bibfnamefont {M.}~\bibnamefont
  {{Nieto-Vesperinas}}},\ }\href {\doibase 10.1103/PhysRevB.61.14119}
  {\bibfield  {journal} {\bibinfo  {journal} {Physical Review B}\ }\textbf
  {\bibinfo {volume} {61}},\ \bibinfo {pages} {14119} (\bibinfo {year}
  {2000})}\BibitemShut {NoStop}%
\bibitem [{\citenamefont {Baffou}\ and\ \citenamefont
  {Quidant}(2013)}]{baffouThermoplasmonicsUsingMetallic2013}%
  \BibitemOpen
  \bibfield  {author} {\bibinfo {author} {\bibfnamefont {G.}~\bibnamefont
  {Baffou}}\ and\ \bibinfo {author} {\bibfnamefont {R.}~\bibnamefont
  {Quidant}},\ }\href {\doibase 10.1002/lpor.201200003} {\bibfield  {journal}
  {\bibinfo  {journal} {Laser \& Photonics Reviews}\ }\textbf {\bibinfo
  {volume} {7}},\ \bibinfo {pages} {171} (\bibinfo {year} {2013})}\BibitemShut
  {NoStop}%
\bibitem [{\citenamefont {Girard}\ \emph {et~al.}(2018)\citenamefont {Girard},
  \citenamefont {Wiecha}, \citenamefont {Cuche},\ and\ \citenamefont
  {Dujardin}}]{girardDesigningThermoplasmonicProperties2018}%
  \BibitemOpen
  \bibfield  {author} {\bibinfo {author} {\bibfnamefont {C.}~\bibnamefont
  {Girard}}, \bibinfo {author} {\bibfnamefont {P.~R.}\ \bibnamefont {Wiecha}},
  \bibinfo {author} {\bibfnamefont {A.}~\bibnamefont {Cuche}}, \ and\ \bibinfo
  {author} {\bibfnamefont {E.}~\bibnamefont {Dujardin}},\ }\href {\doibase
  10.1088/2040-8986/aac934} {\bibfield  {journal} {\bibinfo  {journal} {Journal
  of Optics}\ }\textbf {\bibinfo {volume} {20}},\ \bibinfo {pages} {075004}
  (\bibinfo {year} {2018})}\BibitemShut {NoStop}%
\bibitem [{\citenamefont {Mulholland}\ \emph {et~al.}(1994)\citenamefont
  {Mulholland}, \citenamefont {Bohren},\ and\ \citenamefont
  {Fuller}}]{mulhollandLightScatteringAgglomerates1994}%
  \BibitemOpen
  \bibfield  {author} {\bibinfo {author} {\bibfnamefont {G.~W.}\ \bibnamefont
  {Mulholland}}, \bibinfo {author} {\bibfnamefont {C.~F.}\ \bibnamefont
  {Bohren}}, \ and\ \bibinfo {author} {\bibfnamefont {K.~A.}\ \bibnamefont
  {Fuller}},\ }\href {\doibase 10.1021/la00020a009} {\bibfield  {journal}
  {\bibinfo  {journal} {Langmuir}\ }\textbf {\bibinfo {volume} {10}},\ \bibinfo
  {pages} {2533} (\bibinfo {year} {1994})}\BibitemShut {NoStop}%
\bibitem [{\citenamefont {Huntemann}\ \emph {et~al.}(2011)\citenamefont
  {Huntemann}, \citenamefont {Heygster},\ and\ \citenamefont
  {Hong}}]{huntemannDiscreteDipoleApproximation2011}%
  \BibitemOpen
  \bibfield  {author} {\bibinfo {author} {\bibfnamefont {M.}~\bibnamefont
  {Huntemann}}, \bibinfo {author} {\bibfnamefont {G.}~\bibnamefont {Heygster}},
  \ and\ \bibinfo {author} {\bibfnamefont {G.}~\bibnamefont {Hong}},\ }\href
  {\doibase 10.1016/j.jocs.2011.05.011} {\bibfield  {journal} {\bibinfo
  {journal} {Journal of Computational Science}\ }\bibinfo {series} {Social
  {{Computational Systems}}},\ \textbf {\bibinfo {volume} {2}},\ \bibinfo
  {pages} {262} (\bibinfo {year} {2011})}\BibitemShut {NoStop}%
\bibitem [{\citenamefont
  {Draine}(1988)}]{draineDiscreteDipoleApproximationIts1988}%
  \BibitemOpen
  \bibfield  {author} {\bibinfo {author} {\bibfnamefont {B.~T.}\ \bibnamefont
  {Draine}},\ }\href@noop {} {\bibfield  {journal} {\bibinfo  {journal}
  {Astrophysical Journal}\ }\textbf {\bibinfo {volume} {333}},\ \bibinfo
  {pages} {848} (\bibinfo {year} {1988})}\BibitemShut {NoStop}%
\bibitem [{\citenamefont {Cherukuri}\ \emph {et~al.}(2010)\citenamefont
  {Cherukuri}, \citenamefont {Glazer},\ and\ \citenamefont
  {Curley}}]{cherukuriTargetedHyperthermiaUsing2010}%
  \BibitemOpen
  \bibfield  {author} {\bibinfo {author} {\bibfnamefont {P.}~\bibnamefont
  {Cherukuri}}, \bibinfo {author} {\bibfnamefont {E.~S.}\ \bibnamefont
  {Glazer}}, \ and\ \bibinfo {author} {\bibfnamefont {S.~A.}\ \bibnamefont
  {Curley}},\ }\href {\doibase 10.1016/j.addr.2009.11.006} {\bibfield
  {journal} {\bibinfo  {journal} {Advanced Drug Delivery Reviews}\ }\bibinfo
  {series} {Targeted {{Delivery Using Inorganic Nanosystem}}},\ \textbf
  {\bibinfo {volume} {62}},\ \bibinfo {pages} {339} (\bibinfo {year}
  {2010})}\BibitemShut {NoStop}%
\bibitem [{\citenamefont
  {Stockman}(2011)}]{stockmanNanoplasmonicsPhysicsApplications2011}%
  \BibitemOpen
  \bibfield  {author} {\bibinfo {author} {\bibfnamefont {M.~I.}\ \bibnamefont
  {Stockman}},\ }\href {\doibase 10.1063/1.3554315} {\bibfield  {journal}
  {\bibinfo  {journal} {Physics Today}\ }\textbf {\bibinfo {volume} {64}},\
  \bibinfo {pages} {39} (\bibinfo {year} {2011})}\BibitemShut {NoStop}%
\bibitem [{\citenamefont {Genevet}\ \emph {et~al.}(2017)\citenamefont
  {Genevet}, \citenamefont {Capasso}, \citenamefont {Aieta}, \citenamefont
  {Khorasaninejad},\ and\ \citenamefont
  {Devlin}}]{genevetRecentAdvancesPlanar2017}%
  \BibitemOpen
  \bibfield  {author} {\bibinfo {author} {\bibfnamefont {P.}~\bibnamefont
  {Genevet}}, \bibinfo {author} {\bibfnamefont {F.}~\bibnamefont {Capasso}},
  \bibinfo {author} {\bibfnamefont {F.}~\bibnamefont {Aieta}}, \bibinfo
  {author} {\bibfnamefont {M.}~\bibnamefont {Khorasaninejad}}, \ and\ \bibinfo
  {author} {\bibfnamefont {R.}~\bibnamefont {Devlin}},\ }\href {\doibase
  10.1364/OPTICA.4.000139} {\bibfield  {journal} {\bibinfo  {journal} {Optica}\
  }\textbf {\bibinfo {volume} {4}},\ \bibinfo {pages} {139} (\bibinfo {year}
  {2017})}\BibitemShut {NoStop}%
\bibitem [{\citenamefont {Bai}\ \emph {et~al.}(2013)\citenamefont {Bai},
  \citenamefont {Perrin}, \citenamefont {Sauvan}, \citenamefont {Hugonin},\
  and\ \citenamefont {Lalanne}}]{baiEfficientIntuitiveMethod2013}%
  \BibitemOpen
  \bibfield  {author} {\bibinfo {author} {\bibfnamefont {Q.}~\bibnamefont
  {Bai}}, \bibinfo {author} {\bibfnamefont {M.}~\bibnamefont {Perrin}},
  \bibinfo {author} {\bibfnamefont {C.}~\bibnamefont {Sauvan}}, \bibinfo
  {author} {\bibfnamefont {J.-P.}\ \bibnamefont {Hugonin}}, \ and\ \bibinfo
  {author} {\bibfnamefont {P.}~\bibnamefont {Lalanne}},\ }\href {\doibase
  10.1364/OE.21.027371} {\bibfield  {journal} {\bibinfo  {journal} {Optics
  Express}\ }\textbf {\bibinfo {volume} {21}},\ \bibinfo {pages} {27371}
  (\bibinfo {year} {2013})}\BibitemShut {NoStop}%
\bibitem [{\citenamefont {Lalanne}\ \emph {et~al.}(2018)\citenamefont
  {Lalanne}, \citenamefont {Yan}, \citenamefont {Vynck}, \citenamefont
  {Sauvan},\ and\ \citenamefont
  {Hugonin}}]{lalanneLightInteractionPhotonic2018}%
  \BibitemOpen
  \bibfield  {author} {\bibinfo {author} {\bibfnamefont {P.}~\bibnamefont
  {Lalanne}}, \bibinfo {author} {\bibfnamefont {W.}~\bibnamefont {Yan}},
  \bibinfo {author} {\bibfnamefont {K.}~\bibnamefont {Vynck}}, \bibinfo
  {author} {\bibfnamefont {C.}~\bibnamefont {Sauvan}}, \ and\ \bibinfo {author}
  {\bibfnamefont {J.-P.}\ \bibnamefont {Hugonin}},\ }\href {\doibase
  10.1002/lpor.201700113} {\bibfield  {journal} {\bibinfo  {journal} {Laser \&
  Photonics Reviews}\ }\textbf {\bibinfo {volume} {12}},\ \bibinfo {pages}
  {1700113} (\bibinfo {year} {2018})}\BibitemShut {NoStop}%
\bibitem [{\citenamefont {Kristensen}\ \emph {et~al.}(2015)\citenamefont
  {Kristensen}, \citenamefont {Ge},\ and\ \citenamefont
  {Hughes}}]{kristensenNormalizationQuasinormalModes2015}%
  \BibitemOpen
  \bibfield  {author} {\bibinfo {author} {\bibfnamefont {P.~T.}\ \bibnamefont
  {Kristensen}}, \bibinfo {author} {\bibfnamefont {R.-C.}\ \bibnamefont {Ge}},
  \ and\ \bibinfo {author} {\bibfnamefont {S.}~\bibnamefont {Hughes}},\ }\href
  {\doibase 10.1103/PhysRevA.92.053810} {\bibfield  {journal} {\bibinfo
  {journal} {Physical Review A}\ }\textbf {\bibinfo {volume} {92}},\ \bibinfo
  {pages} {053810} (\bibinfo {year} {2015})}\BibitemShut {NoStop}%
\bibitem [{\citenamefont {Chen}\ \emph {et~al.}(2019)\citenamefont {Chen},
  \citenamefont {Bergman},\ and\ \citenamefont
  {Sivan}}]{chenGeneralizingNormalMode2019}%
  \BibitemOpen
  \bibfield  {author} {\bibinfo {author} {\bibfnamefont {P.~Y.}\ \bibnamefont
  {Chen}}, \bibinfo {author} {\bibfnamefont {D.~J.}\ \bibnamefont {Bergman}}, \
  and\ \bibinfo {author} {\bibfnamefont {Y.}~\bibnamefont {Sivan}},\ }\href
  {\doibase 10.1103/PhysRevApplied.11.044018} {\bibfield  {journal} {\bibinfo
  {journal} {Physical Review Applied}\ }\textbf {\bibinfo {volume} {11}},\
  \bibinfo {pages} {044018} (\bibinfo {year} {2019})}\BibitemShut {NoStop}%
\bibitem [{\citenamefont
  {Jackson}(1999)}]{jacksonClassicalElectrodynamics1999}%
  \BibitemOpen
  \bibfield  {author} {\bibinfo {author} {\bibfnamefont {J.~D.}\ \bibnamefont
  {Jackson}},\ }\href@noop {} {\emph {\bibinfo {title} {Classical
  {{Electrodynamics}}}}},\ \bibinfo {edition} {3rd}\ ed.\ (\bibinfo
  {publisher} {{Wiley}},\ \bibinfo {year} {1999})\BibitemShut {NoStop}%
\bibitem [{\citenamefont {Alaee}\ \emph {et~al.}(2018)\citenamefont {Alaee},
  \citenamefont {Rockstuhl},\ and\ \citenamefont
  {{Fernandez-Corbaton}}}]{alaeeElectromagneticMultipoleExpansion2018}%
  \BibitemOpen
  \bibfield  {author} {\bibinfo {author} {\bibfnamefont {R.}~\bibnamefont
  {Alaee}}, \bibinfo {author} {\bibfnamefont {C.}~\bibnamefont {Rockstuhl}}, \
  and\ \bibinfo {author} {\bibfnamefont {I.}~\bibnamefont
  {{Fernandez-Corbaton}}},\ }\href {\doibase 10.1016/j.optcom.2017.08.064}
  {\bibfield  {journal} {\bibinfo  {journal} {Optics Communications}\ }\textbf
  {\bibinfo {volume} {407}},\ \bibinfo {pages} {17} (\bibinfo {year} {2018})},\
  \Eprint {http://arxiv.org/abs/1701.00755} {arXiv:1701.00755} \BibitemShut
  {NoStop}%
\bibitem [{\citenamefont {Alaee}\ \emph {et~al.}(2019)\citenamefont {Alaee},
  \citenamefont {Rockstuhl},\ and\ \citenamefont
  {{Fernandez-Corbaton}}}]{alaeeExactMultipolarDecompositions2019}%
  \BibitemOpen
  \bibfield  {author} {\bibinfo {author} {\bibfnamefont {R.}~\bibnamefont
  {Alaee}}, \bibinfo {author} {\bibfnamefont {C.}~\bibnamefont {Rockstuhl}}, \
  and\ \bibinfo {author} {\bibfnamefont {I.}~\bibnamefont
  {{Fernandez-Corbaton}}},\ }\href {\doibase 10.1002/adom.201800783} {\bibfield
   {journal} {\bibinfo  {journal} {Advanced Optical Materials}\ }\textbf
  {\bibinfo {volume} {7}},\ \bibinfo {pages} {1800783} (\bibinfo {year}
  {2019})}\BibitemShut {NoStop}%
\bibitem [{\citenamefont {Evlyukhin}\ and\ \citenamefont
  {Chichkov}(2019)}]{evlyukhinMultipoleDecompositionsDirectional2019}%
  \BibitemOpen
  \bibfield  {author} {\bibinfo {author} {\bibfnamefont {A.~B.}\ \bibnamefont
  {Evlyukhin}}\ and\ \bibinfo {author} {\bibfnamefont {B.~N.}\ \bibnamefont
  {Chichkov}},\ }\href {\doibase 10.1103/PhysRevB.100.125415} {\bibfield
  {journal} {\bibinfo  {journal} {Physical Review B}\ }\textbf {\bibinfo
  {volume} {100}},\ \bibinfo {pages} {125415} (\bibinfo {year}
  {2019})}\BibitemShut {NoStop}%
\bibitem [{\citenamefont {Arango}\ and\ \citenamefont
  {Koenderink}(2013)}]{arangoPolarizabilityTensorRetrieval2013}%
  \BibitemOpen
  \bibfield  {author} {\bibinfo {author} {\bibfnamefont {F.~B.}\ \bibnamefont
  {Arango}}\ and\ \bibinfo {author} {\bibfnamefont {A.~F.}\ \bibnamefont
  {Koenderink}},\ }\href {\doibase 10.1088/1367-2630/15/7/073023} {\bibfield
  {journal} {\bibinfo  {journal} {New Journal of Physics}\ }\textbf {\bibinfo
  {volume} {15}},\ \bibinfo {pages} {073023} (\bibinfo {year}
  {2013})}\BibitemShut {NoStop}%
\bibitem [{\citenamefont {Evlyukhin}\ \emph {et~al.}(2011)\citenamefont
  {Evlyukhin}, \citenamefont {Reinhardt},\ and\ \citenamefont
  {Chichkov}}]{evlyukhinMultipoleLightScattering2011}%
  \BibitemOpen
  \bibfield  {author} {\bibinfo {author} {\bibfnamefont {A.~B.}\ \bibnamefont
  {Evlyukhin}}, \bibinfo {author} {\bibfnamefont {C.}~\bibnamefont
  {Reinhardt}}, \ and\ \bibinfo {author} {\bibfnamefont {B.~N.}\ \bibnamefont
  {Chichkov}},\ }\href {\doibase 10.1103/PhysRevB.84.235429} {\bibfield
  {journal} {\bibinfo  {journal} {Physical Review B}\ }\textbf {\bibinfo
  {volume} {84}},\ \bibinfo {pages} {235429} (\bibinfo {year}
  {2011})}\BibitemShut {NoStop}%
\bibitem [{\citenamefont {Hinamoto}\ \emph {et~al.}(2021)\citenamefont
  {Hinamoto}, \citenamefont {Hinamoto}, \citenamefont {Fujii},\ and\
  \citenamefont {Fujii}}]{hinamotoMENPOpensourceMATLAB2021}%
  \BibitemOpen
  \bibfield  {author} {\bibinfo {author} {\bibfnamefont {T.}~\bibnamefont
  {Hinamoto}}, \bibinfo {author} {\bibfnamefont {T.}~\bibnamefont {Hinamoto}},
  \bibinfo {author} {\bibfnamefont {M.}~\bibnamefont {Fujii}}, \ and\ \bibinfo
  {author} {\bibfnamefont {M.}~\bibnamefont {Fujii}},\ }\href {\doibase
  10.1364/OSAC.425189} {\bibfield  {journal} {\bibinfo  {journal} {OSA
  Continuum}\ }\textbf {\bibinfo {volume} {4}},\ \bibinfo {pages} {1640}
  (\bibinfo {year} {2021})}\BibitemShut {NoStop}%
\bibitem [{\citenamefont {Mun}\ \emph {et~al.}(2020)\citenamefont {Mun},
  \citenamefont {So}, \citenamefont {Jang},\ and\ \citenamefont
  {Rho}}]{munDescribingMetaAtomsUsing2020}%
  \BibitemOpen
  \bibfield  {author} {\bibinfo {author} {\bibfnamefont {J.}~\bibnamefont
  {Mun}}, \bibinfo {author} {\bibfnamefont {S.}~\bibnamefont {So}}, \bibinfo
  {author} {\bibfnamefont {J.}~\bibnamefont {Jang}}, \ and\ \bibinfo {author}
  {\bibfnamefont {J.}~\bibnamefont {Rho}},\ }\href {\doibase
  10.1021/acsphotonics.9b01776} {\bibfield  {journal} {\bibinfo  {journal} {ACS
  Photonics}\ }\textbf {\bibinfo {volume} {7}},\ \bibinfo {pages} {1153}
  (\bibinfo {year} {2020})}\BibitemShut {NoStop}%
\bibitem [{\citenamefont {Martin}\ \emph {et~al.}(1995)\citenamefont {Martin},
  \citenamefont {Girard},\ and\ \citenamefont
  {Dereux}}]{martinGeneralizedFieldPropagator1995}%
  \BibitemOpen
  \bibfield  {author} {\bibinfo {author} {\bibfnamefont {O.~J.~F.}\
  \bibnamefont {Martin}}, \bibinfo {author} {\bibfnamefont {C.}~\bibnamefont
  {Girard}}, \ and\ \bibinfo {author} {\bibfnamefont {A.}~\bibnamefont
  {Dereux}},\ }\href {\doibase 10.1103/PhysRevLett.74.526} {\bibfield
  {journal} {\bibinfo  {journal} {Physical Review Letters}\ }\textbf {\bibinfo
  {volume} {74}},\ \bibinfo {pages} {526} (\bibinfo {year} {1995})}\BibitemShut
  {NoStop}%
\bibitem [{\citenamefont {Sersic}\ \emph {et~al.}(2011)\citenamefont {Sersic},
  \citenamefont {Tuambilangana}, \citenamefont {Kampfrath},\ and\ \citenamefont
  {Koenderink}}]{sersicMagnetoelectricPointScattering2011}%
  \BibitemOpen
  \bibfield  {author} {\bibinfo {author} {\bibfnamefont {I.}~\bibnamefont
  {Sersic}}, \bibinfo {author} {\bibfnamefont {C.}~\bibnamefont
  {Tuambilangana}}, \bibinfo {author} {\bibfnamefont {T.}~\bibnamefont
  {Kampfrath}}, \ and\ \bibinfo {author} {\bibfnamefont {A.~F.}\ \bibnamefont
  {Koenderink}},\ }\href {\doibase 10.1103/PhysRevB.83.245102} {\bibfield
  {journal} {\bibinfo  {journal} {Physical Review B}\ }\textbf {\bibinfo
  {volume} {83}},\ \bibinfo {pages} {245102} (\bibinfo {year}
  {2011})}\BibitemShut {NoStop}%
\bibitem [{\citenamefont {Wu}\ \emph {et~al.}(2020)\citenamefont {Wu},
  \citenamefont {Baron}, \citenamefont {Lalanne},\ and\ \citenamefont
  {Vynck}}]{wuIntrinsicMultipolarContents2020}%
  \BibitemOpen
  \bibfield  {author} {\bibinfo {author} {\bibfnamefont {T.}~\bibnamefont
  {Wu}}, \bibinfo {author} {\bibfnamefont {A.}~\bibnamefont {Baron}}, \bibinfo
  {author} {\bibfnamefont {P.}~\bibnamefont {Lalanne}}, \ and\ \bibinfo
  {author} {\bibfnamefont {K.}~\bibnamefont {Vynck}},\ }\href {\doibase
  10.1103/PhysRevA.101.011803} {\bibfield  {journal} {\bibinfo  {journal}
  {Physical Review A}\ }\textbf {\bibinfo {volume} {101}},\ \bibinfo {pages}
  {011803} (\bibinfo {year} {2020})}\BibitemShut {NoStop}%
\bibitem [{\citenamefont {Lunnemann}\ and\ \citenamefont
  {Koenderink}(2016)}]{lunnemannLocalDensityOptical2016}%
  \BibitemOpen
  \bibfield  {author} {\bibinfo {author} {\bibfnamefont {P.}~\bibnamefont
  {Lunnemann}}\ and\ \bibinfo {author} {\bibfnamefont {A.~F.}\ \bibnamefont
  {Koenderink}},\ }\href {\doibase 10.1038/srep20655} {\bibfield  {journal}
  {\bibinfo  {journal} {Scientific Reports}\ }\textbf {\bibinfo {volume} {6}},\
  \bibinfo {pages} {srep20655} (\bibinfo {year} {2016})}\BibitemShut {NoStop}%
\bibitem [{\citenamefont {Patoux}\ \emph {et~al.}(2020)\citenamefont {Patoux},
  \citenamefont {Majorel}, \citenamefont {Wiecha}, \citenamefont {Cuche},
  \citenamefont {Muskens}, \citenamefont {Girard},\ and\ \citenamefont
  {Arbouet}}]{patouxPolarizabilitiesComplexIndividual2020}%
  \BibitemOpen
  \bibfield  {author} {\bibinfo {author} {\bibfnamefont {A.}~\bibnamefont
  {Patoux}}, \bibinfo {author} {\bibfnamefont {C.}~\bibnamefont {Majorel}},
  \bibinfo {author} {\bibfnamefont {P.~R.}\ \bibnamefont {Wiecha}}, \bibinfo
  {author} {\bibfnamefont {A.}~\bibnamefont {Cuche}}, \bibinfo {author}
  {\bibfnamefont {O.~L.}\ \bibnamefont {Muskens}}, \bibinfo {author}
  {\bibfnamefont {C.}~\bibnamefont {Girard}}, \ and\ \bibinfo {author}
  {\bibfnamefont {A.}~\bibnamefont {Arbouet}},\ }\href {\doibase
  10.1103/PhysRevB.101.235418} {\bibfield  {journal} {\bibinfo  {journal}
  {Physical Review B}\ }\textbf {\bibinfo {volume} {101}},\ \bibinfo {pages}
  {235418} (\bibinfo {year} {2020})},\ \Eprint
  {http://arxiv.org/abs/1912.04124} {arXiv:1912.04124} \BibitemShut {NoStop}%
\bibitem [{\citenamefont {Abujetas}\ \emph {et~al.}(2020)\citenamefont
  {Abujetas}, \citenamefont {{Olmos-Trigo}}, \citenamefont {S{\'a}enz},\ and\
  \citenamefont {{S{\'a}nchez-Gil}}}]{abujetasCoupledElectricMagnetic2020}%
  \BibitemOpen
  \bibfield  {author} {\bibinfo {author} {\bibfnamefont {D.~R.}\ \bibnamefont
  {Abujetas}}, \bibinfo {author} {\bibfnamefont {J.}~\bibnamefont
  {{Olmos-Trigo}}}, \bibinfo {author} {\bibfnamefont {J.~J.}\ \bibnamefont
  {S{\'a}enz}}, \ and\ \bibinfo {author} {\bibfnamefont {J.~A.}\ \bibnamefont
  {{S{\'a}nchez-Gil}}},\ }\href {\doibase 10.1103/PhysRevB.102.125411}
  {\bibfield  {journal} {\bibinfo  {journal} {Physical Review B}\ }\textbf
  {\bibinfo {volume} {102}},\ \bibinfo {pages} {125411} (\bibinfo {year}
  {2020})}\BibitemShut {NoStop}%
\bibitem [{\citenamefont
  {Buckingham}(1967)}]{buckinghamPermanentInducedMolecular1967}%
  \BibitemOpen
  \bibfield  {author} {\bibinfo {author} {\bibfnamefont {A.~D.}\ \bibnamefont
  {Buckingham}},\ }\href {\doibase 10.1002/9780470143582.ch2} {\bibfield
  {journal} {\bibinfo  {journal} {Advances in Chemical Physics: Intermolecular
  Forces}\ }\textbf {\bibinfo {volume} {12}},\ \bibinfo {pages} {107} (\bibinfo
  {year} {1967})}\BibitemShut {NoStop}%
\bibitem [{\citenamefont {Baranov}\ \emph {et~al.}(2017)\citenamefont
  {Baranov}, \citenamefont {Savelev}, \citenamefont {Li}, \citenamefont
  {Krasnok},\ and\ \citenamefont
  {Al{\`u}}}]{baranovModifyingMagneticDipole2017}%
  \BibitemOpen
  \bibfield  {author} {\bibinfo {author} {\bibfnamefont {D.~G.}\ \bibnamefont
  {Baranov}}, \bibinfo {author} {\bibfnamefont {R.~S.}\ \bibnamefont
  {Savelev}}, \bibinfo {author} {\bibfnamefont {S.~V.}\ \bibnamefont {Li}},
  \bibinfo {author} {\bibfnamefont {A.~E.}\ \bibnamefont {Krasnok}}, \ and\
  \bibinfo {author} {\bibfnamefont {A.}~\bibnamefont {Al{\`u}}},\ }\href
  {\doibase 10.1002/lpor.201600268} {\bibfield  {journal} {\bibinfo  {journal}
  {Laser \& Photonics Reviews}\ }\textbf {\bibinfo {volume} {11}},\ \bibinfo
  {pages} {1600268} (\bibinfo {year} {2017})}\BibitemShut {NoStop}%
\bibitem [{\citenamefont {Wiecha}\ \emph {et~al.}(2018)\citenamefont {Wiecha},
  \citenamefont {Arbouet}, \citenamefont {Cuche}, \citenamefont {Paillard},\
  and\ \citenamefont {Girard}}]{wiechaDecayRateMagnetic2018}%
  \BibitemOpen
  \bibfield  {author} {\bibinfo {author} {\bibfnamefont {P.~R.}\ \bibnamefont
  {Wiecha}}, \bibinfo {author} {\bibfnamefont {A.}~\bibnamefont {Arbouet}},
  \bibinfo {author} {\bibfnamefont {A.}~\bibnamefont {Cuche}}, \bibinfo
  {author} {\bibfnamefont {V.}~\bibnamefont {Paillard}}, \ and\ \bibinfo
  {author} {\bibfnamefont {C.}~\bibnamefont {Girard}},\ }\href {\doibase
  10.1103/PhysRevB.97.085411} {\bibfield  {journal} {\bibinfo  {journal}
  {Physical Review B}\ }\textbf {\bibinfo {volume} {97}},\ \bibinfo {pages}
  {085411} (\bibinfo {year} {2018})}\BibitemShut {NoStop}%
\bibitem [{\citenamefont {Girard}(2005)}]{girardFieldsNanostructures2005}%
  \BibitemOpen
  \bibfield  {author} {\bibinfo {author} {\bibfnamefont {C.}~\bibnamefont
  {Girard}},\ }\href {\doibase 10.1088/0034-4885/68/8/R05} {\bibfield
  {journal} {\bibinfo  {journal} {Reports on Progress in Physics}\ }\textbf
  {\bibinfo {volume} {68}},\ \bibinfo {pages} {1883} (\bibinfo {year}
  {2005})}\BibitemShut {NoStop}%
\bibitem [{\citenamefont {Novotny}\ and\ \citenamefont
  {Hecht}(2006)}]{novotnyPrinciplesNanooptics2006}%
  \BibitemOpen
  \bibfield  {author} {\bibinfo {author} {\bibfnamefont {L.}~\bibnamefont
  {Novotny}}\ and\ \bibinfo {author} {\bibfnamefont {B.}~\bibnamefont
  {Hecht}},\ }\href@noop {} {\emph {\bibinfo {title} {Principles of
  Nano-Optics}}}\ (\bibinfo  {publisher} {{Cambridge University Press}},\
  \bibinfo {address} {{Cambridge ; New York}},\ \bibinfo {year}
  {2006})\BibitemShut {NoStop}%
\bibitem [{\citenamefont {Girard}\ \emph {et~al.}(2008)\citenamefont {Girard},
  \citenamefont {Dujardin}, \citenamefont {Baffou},\ and\ \citenamefont
  {Quidant}}]{girardShapingManipulationLight2008}%
  \BibitemOpen
  \bibfield  {author} {\bibinfo {author} {\bibfnamefont {C.}~\bibnamefont
  {Girard}}, \bibinfo {author} {\bibfnamefont {E.}~\bibnamefont {Dujardin}},
  \bibinfo {author} {\bibfnamefont {G.}~\bibnamefont {Baffou}}, \ and\ \bibinfo
  {author} {\bibfnamefont {R.}~\bibnamefont {Quidant}},\ }\href {\doibase
  10.1088/1367-2630/10/10/105016} {\bibfield  {journal} {\bibinfo  {journal}
  {New Journal of Physics}\ }\textbf {\bibinfo {volume} {10}},\ \bibinfo
  {pages} {105016} (\bibinfo {year} {2008})}\BibitemShut {NoStop}%
\bibitem [{\citenamefont {Wiecha}(2018)}]{wiechaPyGDMPythonToolkit2018}%
  \BibitemOpen
  \bibfield  {author} {\bibinfo {author} {\bibfnamefont {P.~R.}\ \bibnamefont
  {Wiecha}},\ }\href {\doibase 10.1016/j.cpc.2018.06.017} {\bibfield  {journal}
  {\bibinfo  {journal} {Computer Physics Communications}\ }\textbf {\bibinfo
  {volume} {233}},\ \bibinfo {pages} {167} (\bibinfo {year}
  {2018})}\BibitemShut {NoStop}%
\bibitem [{\citenamefont {Wiecha}\ \emph {et~al.}(2022)\citenamefont {Wiecha},
  \citenamefont {Majorel}, \citenamefont {Arbouet}, \citenamefont {Patoux},
  \citenamefont {Br{\^u}l{\'e}}, \citenamefont {des Francs},\ and\
  \citenamefont {Girard}}]{wiechaPyGDMNewFunctionalities2022}%
  \BibitemOpen
  \bibfield  {author} {\bibinfo {author} {\bibfnamefont {P.~R.}\ \bibnamefont
  {Wiecha}}, \bibinfo {author} {\bibfnamefont {C.}~\bibnamefont {Majorel}},
  \bibinfo {author} {\bibfnamefont {A.}~\bibnamefont {Arbouet}}, \bibinfo
  {author} {\bibfnamefont {A.}~\bibnamefont {Patoux}}, \bibinfo {author}
  {\bibfnamefont {Y.}~\bibnamefont {Br{\^u}l{\'e}}}, \bibinfo {author}
  {\bibfnamefont {G.~C.}\ \bibnamefont {des Francs}}, \ and\ \bibinfo {author}
  {\bibfnamefont {C.}~\bibnamefont {Girard}},\ }\href {\doibase
  10.1016/j.cpc.2021.108142} {\bibfield  {journal} {\bibinfo  {journal}
  {Computer Physics Communications}\ }\textbf {\bibinfo {volume} {270}},\
  \bibinfo {pages} {108142} (\bibinfo {year} {2022})},\ \Eprint
  {http://arxiv.org/abs/2105.04587} {arXiv:2105.04587} \BibitemShut {NoStop}%
\bibitem [{\citenamefont {Dubovik}\ and\ \citenamefont
  {Tugushev}(1990)}]{dubovikToroidMomentsElectrodynamics1990}%
  \BibitemOpen
  \bibfield  {author} {\bibinfo {author} {\bibfnamefont {V.~M.}\ \bibnamefont
  {Dubovik}}\ and\ \bibinfo {author} {\bibfnamefont {V.~V.}\ \bibnamefont
  {Tugushev}},\ }\href {\doibase 10.1016/0370-1573(90)90042-Z} {\bibfield
  {journal} {\bibinfo  {journal} {Physics Reports}\ }\textbf {\bibinfo {volume}
  {187}},\ \bibinfo {pages} {145} (\bibinfo {year} {1990})}\BibitemShut
  {NoStop}%
\bibitem [{\citenamefont
  {Waterman}(1965)}]{watermanMatrixFormulationElectromagnetic1965}%
  \BibitemOpen
  \bibfield  {author} {\bibinfo {author} {\bibfnamefont {P.}~\bibnamefont
  {Waterman}},\ }\href {\doibase 10.1109/PROC.1965.4058} {\bibfield  {journal}
  {\bibinfo  {journal} {Proceedings of the IEEE}\ }\textbf {\bibinfo {volume}
  {53}},\ \bibinfo {pages} {805} (\bibinfo {year} {1965})}\BibitemShut
  {NoStop}%
\bibitem [{\citenamefont {Mishchenko}\ \emph {et~al.}(2002)\citenamefont
  {Mishchenko}, \citenamefont {Travis},\ and\ \citenamefont
  {Lacis}}]{mishchenkoScatteringAbsorptionEmission2002}%
  \BibitemOpen
  \bibfield  {author} {\bibinfo {author} {\bibfnamefont {M.~I.}\ \bibnamefont
  {Mishchenko}}, \bibinfo {author} {\bibfnamefont {L.~D.}\ \bibnamefont
  {Travis}}, \ and\ \bibinfo {author} {\bibfnamefont {A.~A.}\ \bibnamefont
  {Lacis}},\ }\href@noop {} {\emph {\bibinfo {title} {Scattering,
  {{Absorption}}, and {{Emission}} of {{Light}} by {{Small Particles}}}}}\
  (\bibinfo  {publisher} {{Cambridge University Press}},\ \bibinfo {address}
  {{Cambridge}},\ \bibinfo {year} {2002})\BibitemShut {NoStop}%
\bibitem [{\citenamefont
  {Mishchenko}(2008)}]{mishchenkoMultipleScatteringRadiative2008}%
  \BibitemOpen
  \bibfield  {author} {\bibinfo {author} {\bibfnamefont {M.~I.}\ \bibnamefont
  {Mishchenko}},\ }\href {\doibase 10.1029/2007RG000230} {\bibfield  {journal}
  {\bibinfo  {journal} {Reviews of Geophysics}\ }\textbf {\bibinfo {volume}
  {46}} (\bibinfo {year} {2008}),\ 10.1029/2007RG000230}\BibitemShut {NoStop}%
\bibitem [{\citenamefont
  {Litvinov}(2008)}]{litvinovDerivationExtendedBoundary2008}%
  \BibitemOpen
  \bibfield  {author} {\bibinfo {author} {\bibfnamefont {P.}~\bibnamefont
  {Litvinov}},\ }\href {\doibase 10.1016/j.jqsrt.2007.11.011} {\bibfield
  {journal} {\bibinfo  {journal} {Journal of Quantitative Spectroscopy and
  Radiative Transfer}\ }\bibinfo {series} {X {{Conference}} on
  {{Electromagnetic}} and {{Light Scattering}} by {{Non-Spherical
  Particles}}},\ \textbf {\bibinfo {volume} {109}},\ \bibinfo {pages} {1440}
  (\bibinfo {year} {2008})}\BibitemShut {NoStop}%
\bibitem [{\citenamefont {Loke}\ \emph {et~al.}(2009)\citenamefont {Loke},
  \citenamefont {Nieminen}, \citenamefont {Heckenberg},\ and\ \citenamefont
  {{Rubinsztein-Dunlop}}}]{lokeTmatrixCalculationDiscrete2009}%
  \BibitemOpen
  \bibfield  {author} {\bibinfo {author} {\bibfnamefont {V.~L.~Y.}\
  \bibnamefont {Loke}}, \bibinfo {author} {\bibfnamefont {T.~A.}\ \bibnamefont
  {Nieminen}}, \bibinfo {author} {\bibfnamefont {N.~R.}\ \bibnamefont
  {Heckenberg}}, \ and\ \bibinfo {author} {\bibfnamefont {H.}~\bibnamefont
  {{Rubinsztein-Dunlop}}},\ }\href {\doibase 10.1016/j.jqsrt.2009.01.013}
  {\bibfield  {journal} {\bibinfo  {journal} {Journal of Quantitative
  Spectroscopy and Radiative Transfer}\ }\bibinfo {series} {{{XI Conference}}
  on {{Electromagnetic}} and {{Light Scattering}} by {{Non-Spherical
  Particles}}: 2008},\ \textbf {\bibinfo {volume} {110}},\ \bibinfo {pages}
  {1460} (\bibinfo {year} {2009})}\BibitemShut {NoStop}%
\bibitem [{\citenamefont {Fruhnert}\ \emph {et~al.}(2017)\citenamefont
  {Fruhnert}, \citenamefont {{Fernandez-Corbaton}}, \citenamefont
  {Yannopapas},\ and\ \citenamefont
  {Rockstuhl}}]{fruhnertComputingTmatrixScattering2017}%
  \BibitemOpen
  \bibfield  {author} {\bibinfo {author} {\bibfnamefont {M.}~\bibnamefont
  {Fruhnert}}, \bibinfo {author} {\bibfnamefont {I.}~\bibnamefont
  {{Fernandez-Corbaton}}}, \bibinfo {author} {\bibfnamefont {V.}~\bibnamefont
  {Yannopapas}}, \ and\ \bibinfo {author} {\bibfnamefont {C.}~\bibnamefont
  {Rockstuhl}},\ }\href {\doibase 10.3762/bjnano.8.66} {\bibfield  {journal}
  {\bibinfo  {journal} {Beilstein Journal of Nanotechnology}\ }\textbf
  {\bibinfo {volume} {8}},\ \bibinfo {pages} {614} (\bibinfo {year}
  {2017})}\BibitemShut {NoStop}%
\bibitem [{\citenamefont {Bertrand}\ \emph {et~al.}(2020)\citenamefont
  {Bertrand}, \citenamefont {Devilez}, \citenamefont {Hugonin}, \citenamefont
  {Lalanne},\ and\ \citenamefont
  {Vynck}}]{bertrandGlobalPolarizabilityMatrix2020}%
  \BibitemOpen
  \bibfield  {author} {\bibinfo {author} {\bibfnamefont {M.}~\bibnamefont
  {Bertrand}}, \bibinfo {author} {\bibfnamefont {A.}~\bibnamefont {Devilez}},
  \bibinfo {author} {\bibfnamefont {J.-P.}\ \bibnamefont {Hugonin}}, \bibinfo
  {author} {\bibfnamefont {P.}~\bibnamefont {Lalanne}}, \ and\ \bibinfo
  {author} {\bibfnamefont {K.}~\bibnamefont {Vynck}},\ }\href {\doibase
  10.1364/JOSAA.37.000070} {\bibfield  {journal} {\bibinfo  {journal} {JOSA A}\
  }\textbf {\bibinfo {volume} {37}},\ \bibinfo {pages} {70} (\bibinfo {year}
  {2020})},\ \Eprint {http://arxiv.org/abs/1907.12823} {arXiv:1907.12823}
  \BibitemShut {NoStop}%
\bibitem [{\citenamefont {Egel}\ \emph {et~al.}(2016)\citenamefont {Egel},
  \citenamefont {Theobald}, \citenamefont {Donie}, \citenamefont {Lemmer},\
  and\ \citenamefont {Gomard}}]{egelLightScatteringOblate2016}%
  \BibitemOpen
  \bibfield  {author} {\bibinfo {author} {\bibfnamefont {A.}~\bibnamefont
  {Egel}}, \bibinfo {author} {\bibfnamefont {D.}~\bibnamefont {Theobald}},
  \bibinfo {author} {\bibfnamefont {Y.}~\bibnamefont {Donie}}, \bibinfo
  {author} {\bibfnamefont {U.}~\bibnamefont {Lemmer}}, \ and\ \bibinfo {author}
  {\bibfnamefont {G.}~\bibnamefont {Gomard}},\ }\href {\doibase
  10.1364/OE.24.025154} {\bibfield  {journal} {\bibinfo  {journal} {Optics
  Express}\ }\textbf {\bibinfo {volume} {24}},\ \bibinfo {pages} {25154}
  (\bibinfo {year} {2016})}\BibitemShut {NoStop}%
\bibitem [{\citenamefont {Dem{\'e}sy}\ \emph {et~al.}(2018)\citenamefont
  {Dem{\'e}sy}, \citenamefont {Auger},\ and\ \citenamefont
  {Stout}}]{demesyScatteringMatrixArbitrarily2018}%
  \BibitemOpen
  \bibfield  {author} {\bibinfo {author} {\bibfnamefont {G.}~\bibnamefont
  {Dem{\'e}sy}}, \bibinfo {author} {\bibfnamefont {J.-C.}\ \bibnamefont
  {Auger}}, \ and\ \bibinfo {author} {\bibfnamefont {B.}~\bibnamefont
  {Stout}},\ }\href {\doibase 10.1364/JOSAA.35.001401} {\bibfield  {journal}
  {\bibinfo  {journal} {JOSA A}\ }\textbf {\bibinfo {volume} {35}},\ \bibinfo
  {pages} {1401} (\bibinfo {year} {2018})}\BibitemShut {NoStop}%
\bibitem [{\citenamefont {Martin}(2019)}]{martinTmatrixMethodClosely2019}%
  \BibitemOpen
  \bibfield  {author} {\bibinfo {author} {\bibfnamefont {T.}~\bibnamefont
  {Martin}},\ }\href {\doibase 10.1016/j.jqsrt.2019.06.001} {\bibfield
  {journal} {\bibinfo  {journal} {Journal of Quantitative Spectroscopy and
  Radiative Transfer}\ }\textbf {\bibinfo {volume} {234}},\ \bibinfo {pages}
  {40} (\bibinfo {year} {2019})}\BibitemShut {NoStop}%
\bibitem [{\citenamefont {Pattelli}\ \emph {et~al.}(2018)\citenamefont
  {Pattelli}, \citenamefont {Egel}, \citenamefont {Lemmer},\ and\ \citenamefont
  {Wiersma}}]{pattelliRolePackingDensity2018}%
  \BibitemOpen
  \bibfield  {author} {\bibinfo {author} {\bibfnamefont {L.}~\bibnamefont
  {Pattelli}}, \bibinfo {author} {\bibfnamefont {A.}~\bibnamefont {Egel}},
  \bibinfo {author} {\bibfnamefont {U.}~\bibnamefont {Lemmer}}, \ and\ \bibinfo
  {author} {\bibfnamefont {D.~S.}\ \bibnamefont {Wiersma}},\ }\href {\doibase
  10.1364/OPTICA.5.001037} {\bibfield  {journal} {\bibinfo  {journal} {Optica}\
  }\textbf {\bibinfo {volume} {5}},\ \bibinfo {pages} {1037} (\bibinfo {year}
  {2018})}\BibitemShut {NoStop}%
\bibitem [{\citenamefont {Werdehausen}\ \emph {et~al.}(2020)\citenamefont
  {Werdehausen}, \citenamefont {Santiago}, \citenamefont {Burger},
  \citenamefont {Staude}, \citenamefont {Pertsch}, \citenamefont {Rockstuhl},\
  and\ \citenamefont {Decker}}]{werdehausenModelingOpticalMaterials2020}%
  \BibitemOpen
  \bibfield  {author} {\bibinfo {author} {\bibfnamefont {D.}~\bibnamefont
  {Werdehausen}}, \bibinfo {author} {\bibfnamefont {X.~G.}\ \bibnamefont
  {Santiago}}, \bibinfo {author} {\bibfnamefont {S.}~\bibnamefont {Burger}},
  \bibinfo {author} {\bibfnamefont {I.}~\bibnamefont {Staude}}, \bibinfo
  {author} {\bibfnamefont {T.}~\bibnamefont {Pertsch}}, \bibinfo {author}
  {\bibfnamefont {C.}~\bibnamefont {Rockstuhl}}, \ and\ \bibinfo {author}
  {\bibfnamefont {M.}~\bibnamefont {Decker}},\ }\href {\doibase
  10.1002/adts.202000192} {\bibfield  {journal} {\bibinfo  {journal} {Advanced
  Theory and Simulations}\ }\textbf {\bibinfo {volume} {3}},\ \bibinfo {pages}
  {2000192} (\bibinfo {year} {2020})}\BibitemShut {NoStop}%
\bibitem [{\citenamefont {Skarda}\ \emph {et~al.}(2022)\citenamefont {Skarda},
  \citenamefont {Trivedi}, \citenamefont {Su}, \citenamefont {{Ahmad-Stein}},
  \citenamefont {Kwon}, \citenamefont {Han}, \citenamefont {Fan},\ and\
  \citenamefont {Vu{\v
  c}kovi{\'c}}}]{skardaLowoverheadDistributionStrategy2022}%
  \BibitemOpen
  \bibfield  {author} {\bibinfo {author} {\bibfnamefont {J.}~\bibnamefont
  {Skarda}}, \bibinfo {author} {\bibfnamefont {R.}~\bibnamefont {Trivedi}},
  \bibinfo {author} {\bibfnamefont {L.}~\bibnamefont {Su}}, \bibinfo {author}
  {\bibfnamefont {D.}~\bibnamefont {{Ahmad-Stein}}}, \bibinfo {author}
  {\bibfnamefont {H.}~\bibnamefont {Kwon}}, \bibinfo {author} {\bibfnamefont
  {S.}~\bibnamefont {Han}}, \bibinfo {author} {\bibfnamefont {S.}~\bibnamefont
  {Fan}}, \ and\ \bibinfo {author} {\bibfnamefont {J.}~\bibnamefont {Vu{\v
  c}kovi{\'c}}},\ }\href {\doibase 10.1038/s41524-022-00774-y} {\bibfield
  {journal} {\bibinfo  {journal} {npj Computational Materials}\ }\textbf
  {\bibinfo {volume} {8}},\ \bibinfo {pages} {1} (\bibinfo {year}
  {2022})}\BibitemShut {NoStop}%
\bibitem [{\citenamefont {Rahimzadegan}\ \emph {et~al.}(2022)\citenamefont
  {Rahimzadegan}, \citenamefont {Karamanos}, \citenamefont {Alaee},
  \citenamefont {Lamprianidis}, \citenamefont {Beutel}, \citenamefont {Boyd},\
  and\ \citenamefont
  {Rockstuhl}}]{rahimzadeganComprehensiveMultipolarTheory2022}%
  \BibitemOpen
  \bibfield  {author} {\bibinfo {author} {\bibfnamefont {A.}~\bibnamefont
  {Rahimzadegan}}, \bibinfo {author} {\bibfnamefont {T.~D.}\ \bibnamefont
  {Karamanos}}, \bibinfo {author} {\bibfnamefont {R.}~\bibnamefont {Alaee}},
  \bibinfo {author} {\bibfnamefont {A.~G.}\ \bibnamefont {Lamprianidis}},
  \bibinfo {author} {\bibfnamefont {D.}~\bibnamefont {Beutel}}, \bibinfo
  {author} {\bibfnamefont {R.~W.}\ \bibnamefont {Boyd}}, \ and\ \bibinfo
  {author} {\bibfnamefont {C.}~\bibnamefont {Rockstuhl}},\ }\href {\doibase
  10.1002/adom.202102059} {\bibfield  {journal} {\bibinfo  {journal} {Advanced
  Optical Materials}\ }\textbf {\bibinfo {volume} {10}},\ \bibinfo {pages}
  {2102059} (\bibinfo {year} {2022})}\BibitemShut {NoStop}%
\bibitem [{\citenamefont {Miroshnichenko}\ \emph {et~al.}(2015)\citenamefont
  {Miroshnichenko}, \citenamefont {Evlyukhin}, \citenamefont {Yu},
  \citenamefont {Bakker}, \citenamefont {Chipouline}, \citenamefont
  {Kuznetsov}, \citenamefont {Luk'yanchuk}, \citenamefont {Chichkov},\ and\
  \citenamefont {Kivshar}}]{miroshnichenkoNonradiatingAnapoleModes2015}%
  \BibitemOpen
  \bibfield  {author} {\bibinfo {author} {\bibfnamefont {A.~E.}\ \bibnamefont
  {Miroshnichenko}}, \bibinfo {author} {\bibfnamefont {A.~B.}\ \bibnamefont
  {Evlyukhin}}, \bibinfo {author} {\bibfnamefont {Y.~F.}\ \bibnamefont {Yu}},
  \bibinfo {author} {\bibfnamefont {R.~M.}\ \bibnamefont {Bakker}}, \bibinfo
  {author} {\bibfnamefont {A.}~\bibnamefont {Chipouline}}, \bibinfo {author}
  {\bibfnamefont {A.~I.}\ \bibnamefont {Kuznetsov}}, \bibinfo {author}
  {\bibfnamefont {B.}~\bibnamefont {Luk'yanchuk}}, \bibinfo {author}
  {\bibfnamefont {B.~N.}\ \bibnamefont {Chichkov}}, \ and\ \bibinfo {author}
  {\bibfnamefont {Y.~S.}\ \bibnamefont {Kivshar}},\ }\href {\doibase
  10.1038/ncomms9069} {\bibfield  {journal} {\bibinfo  {journal} {Nature
  Communications}\ }\textbf {\bibinfo {volume} {6}},\ \bibinfo {pages} {8069}
  (\bibinfo {year} {2015})}\BibitemShut {NoStop}%
\bibitem [{\citenamefont {Das}\ \emph {et~al.}(2015)\citenamefont {Das},
  \citenamefont {Iyer}, \citenamefont {DeCrescent},\ and\ \citenamefont
  {Schuller}}]{dasBeamEngineeringSelective2015}%
  \BibitemOpen
  \bibfield  {author} {\bibinfo {author} {\bibfnamefont {T.}~\bibnamefont
  {Das}}, \bibinfo {author} {\bibfnamefont {P.~P.}\ \bibnamefont {Iyer}},
  \bibinfo {author} {\bibfnamefont {R.~A.}\ \bibnamefont {DeCrescent}}, \ and\
  \bibinfo {author} {\bibfnamefont {J.~A.}\ \bibnamefont {Schuller}},\ }\href
  {\doibase 10.1103/PhysRevB.92.241110} {\bibfield  {journal} {\bibinfo
  {journal} {Physical Review B}\ }\textbf {\bibinfo {volume} {92}},\ \bibinfo
  {pages} {241110} (\bibinfo {year} {2015})}\BibitemShut {NoStop}%
\bibitem [{\citenamefont {Montagnac}\ \emph {et~al.}(2022)\citenamefont
  {Montagnac}, \citenamefont {Agez}, \citenamefont {Patoux}, \citenamefont
  {Arbouet},\ and\ \citenamefont
  {Paillard}}]{montagnacEngineeredFarfieldOptical2022}%
  \BibitemOpen
  \bibfield  {author} {\bibinfo {author} {\bibfnamefont {M.}~\bibnamefont
  {Montagnac}}, \bibinfo {author} {\bibfnamefont {G.}~\bibnamefont {Agez}},
  \bibinfo {author} {\bibfnamefont {A.}~\bibnamefont {Patoux}}, \bibinfo
  {author} {\bibfnamefont {A.}~\bibnamefont {Arbouet}}, \ and\ \bibinfo
  {author} {\bibfnamefont {V.}~\bibnamefont {Paillard}},\ }\href {\doibase
  10.1063/5.0085940} {\bibfield  {journal} {\bibinfo  {journal} {Journal of
  Applied Physics}\ }\textbf {\bibinfo {volume} {131}},\ \bibinfo {pages}
  {133101} (\bibinfo {year} {2022})},\ \Eprint
  {http://arxiv.org/abs/2107.06058} {arXiv:2107.06058} \BibitemShut {NoStop}%
\bibitem [{\citenamefont {Decker}\ \emph {et~al.}(2015)\citenamefont {Decker},
  \citenamefont {Staude}, \citenamefont {Falkner}, \citenamefont {Dominguez},
  \citenamefont {Neshev}, \citenamefont {Brener}, \citenamefont {Pertsch},\
  and\ \citenamefont {Kivshar}}]{deckerHighEfficiencyDielectricHuygens2015}%
  \BibitemOpen
  \bibfield  {author} {\bibinfo {author} {\bibfnamefont {M.}~\bibnamefont
  {Decker}}, \bibinfo {author} {\bibfnamefont {I.}~\bibnamefont {Staude}},
  \bibinfo {author} {\bibfnamefont {M.}~\bibnamefont {Falkner}}, \bibinfo
  {author} {\bibfnamefont {J.}~\bibnamefont {Dominguez}}, \bibinfo {author}
  {\bibfnamefont {D.~N.}\ \bibnamefont {Neshev}}, \bibinfo {author}
  {\bibfnamefont {I.}~\bibnamefont {Brener}}, \bibinfo {author} {\bibfnamefont
  {T.}~\bibnamefont {Pertsch}}, \ and\ \bibinfo {author} {\bibfnamefont
  {Y.~S.}\ \bibnamefont {Kivshar}},\ }\href {\doibase 10.1002/adom.201400584}
  {\bibfield  {journal} {\bibinfo  {journal} {Advanced Optical Materials}\
  }\textbf {\bibinfo {volume} {3}},\ \bibinfo {pages} {813} (\bibinfo {year}
  {2015})}\BibitemShut {NoStop}%
\bibitem [{\citenamefont {Kerker}\ \emph {et~al.}(1983)\citenamefont {Kerker},
  \citenamefont {Wang},\ and\ \citenamefont
  {Giles}}]{kerkerElectromagneticScatteringMagnetic1983}%
  \BibitemOpen
  \bibfield  {author} {\bibinfo {author} {\bibfnamefont {M.}~\bibnamefont
  {Kerker}}, \bibinfo {author} {\bibfnamefont {D.-S.}\ \bibnamefont {Wang}}, \
  and\ \bibinfo {author} {\bibfnamefont {C.~L.}\ \bibnamefont {Giles}},\ }\href
  {\doibase 10.1364/JOSA.73.000765} {\bibfield  {journal} {\bibinfo  {journal}
  {Journal of the Optical Society of America}\ }\textbf {\bibinfo {volume}
  {73}},\ \bibinfo {pages} {765} (\bibinfo {year} {1983})}\BibitemShut
  {NoStop}%
\bibitem [{\citenamefont {Pfeiffer}\ and\ \citenamefont
  {Grbic}(2013)}]{pfeifferMetamaterialHuygensSurfaces2013}%
  \BibitemOpen
  \bibfield  {author} {\bibinfo {author} {\bibfnamefont {C.}~\bibnamefont
  {Pfeiffer}}\ and\ \bibinfo {author} {\bibfnamefont {A.}~\bibnamefont
  {Grbic}},\ }\href {\doibase 10.1103/PhysRevLett.110.197401} {\bibfield
  {journal} {\bibinfo  {journal} {Physical Review Letters}\ }\textbf {\bibinfo
  {volume} {110}},\ \bibinfo {pages} {197401} (\bibinfo {year}
  {2013})}\BibitemShut {NoStop}%
\bibitem [{\citenamefont {Marco}\ \emph {et~al.}(2021)\citenamefont {Marco},
  \citenamefont {Jiang}, \citenamefont {Fang}, \citenamefont {Lacomme},
  \citenamefont {Zheng}, \citenamefont {Baron}, \citenamefont {Korgel},
  \citenamefont {Barois}, \citenamefont {Drisko},\ and\ \citenamefont
  {Aymonier}}]{marcoBroadbandForwardLight2021}%
  \BibitemOpen
  \bibfield  {author} {\bibinfo {author} {\bibfnamefont {M.~L.~D.}\
  \bibnamefont {Marco}}, \bibinfo {author} {\bibfnamefont {T.}~\bibnamefont
  {Jiang}}, \bibinfo {author} {\bibfnamefont {J.}~\bibnamefont {Fang}},
  \bibinfo {author} {\bibfnamefont {S.}~\bibnamefont {Lacomme}}, \bibinfo
  {author} {\bibfnamefont {Y.}~\bibnamefont {Zheng}}, \bibinfo {author}
  {\bibfnamefont {A.}~\bibnamefont {Baron}}, \bibinfo {author} {\bibfnamefont
  {B.~A.}\ \bibnamefont {Korgel}}, \bibinfo {author} {\bibfnamefont
  {P.}~\bibnamefont {Barois}}, \bibinfo {author} {\bibfnamefont {G.~L.}\
  \bibnamefont {Drisko}}, \ and\ \bibinfo {author} {\bibfnamefont
  {C.}~\bibnamefont {Aymonier}},\ }\href {\doibase 10.1002/adfm.202100915}
  {\bibfield  {journal} {\bibinfo  {journal} {Advanced Functional Materials}\
  ,\ \bibinfo {pages} {2100915}} (\bibinfo {year} {2021})}\BibitemShut
  {NoStop}%
\bibitem [{\citenamefont {Edwards}(1997)}]{edwardsSiliconSi1997}%
  \BibitemOpen
  \bibfield  {author} {\bibinfo {author} {\bibfnamefont {D.~F.}\ \bibnamefont
  {Edwards}},\ }in\ \href@noop {} {\emph {\bibinfo {booktitle} {Handbook of
  {{Optical Constants}} of {{Solids}}}}},\ \bibinfo {editor} {edited by\
  \bibinfo {editor} {\bibfnamefont {E.~D.}\ \bibnamefont {Palik}}}\ (\bibinfo
  {publisher} {{Academic Press}},\ \bibinfo {address} {{Burlington}},\ \bibinfo
  {year} {1997})\ pp.\ \bibinfo {pages} {547--569}\BibitemShut {NoStop}%
\bibitem [{\citenamefont {Manna}\ \emph {et~al.}(2020)\citenamefont {Manna},
  \citenamefont {Sugimoto}, \citenamefont {Eggena}, \citenamefont {Coe},
  \citenamefont {Wang}, \citenamefont {Biswas},\ and\ \citenamefont
  {Fujii}}]{mannaSelectiveExcitationEnhancement2020}%
  \BibitemOpen
  \bibfield  {author} {\bibinfo {author} {\bibfnamefont {U.}~\bibnamefont
  {Manna}}, \bibinfo {author} {\bibfnamefont {H.}~\bibnamefont {Sugimoto}},
  \bibinfo {author} {\bibfnamefont {D.}~\bibnamefont {Eggena}}, \bibinfo
  {author} {\bibfnamefont {B.}~\bibnamefont {Coe}}, \bibinfo {author}
  {\bibfnamefont {R.}~\bibnamefont {Wang}}, \bibinfo {author} {\bibfnamefont
  {M.}~\bibnamefont {Biswas}}, \ and\ \bibinfo {author} {\bibfnamefont
  {M.}~\bibnamefont {Fujii}},\ }\href {\doibase 10.1063/1.5132791} {\bibfield
  {journal} {\bibinfo  {journal} {Journal of Applied Physics}\ }\textbf
  {\bibinfo {volume} {127}},\ \bibinfo {pages} {033101} (\bibinfo {year}
  {2020})}\BibitemShut {NoStop}%
\bibitem [{\citenamefont {Gigli}\ \emph {et~al.}(2021)\citenamefont {Gigli},
  \citenamefont {Li}, \citenamefont {Chavel}, \citenamefont {Leo},
  \citenamefont {Brongersma},\ and\ \citenamefont
  {Lalanne}}]{gigliFundamentalLimitationsHuygens2021}%
  \BibitemOpen
  \bibfield  {author} {\bibinfo {author} {\bibfnamefont {C.}~\bibnamefont
  {Gigli}}, \bibinfo {author} {\bibfnamefont {Q.}~\bibnamefont {Li}}, \bibinfo
  {author} {\bibfnamefont {P.}~\bibnamefont {Chavel}}, \bibinfo {author}
  {\bibfnamefont {G.}~\bibnamefont {Leo}}, \bibinfo {author} {\bibfnamefont
  {M.~L.}\ \bibnamefont {Brongersma}}, \ and\ \bibinfo {author} {\bibfnamefont
  {P.}~\bibnamefont {Lalanne}},\ }\href {\doibase 10.1002/lpor.202000448}
  {\bibfield  {journal} {\bibinfo  {journal} {Laser \& Photonics Reviews}\
  }\textbf {\bibinfo {volume} {15}},\ \bibinfo {pages} {2000448} (\bibinfo
  {year} {2021})}\BibitemShut {NoStop}%
\bibitem [{\citenamefont {Volpe}\ \emph {et~al.}(2009)\citenamefont {Volpe},
  \citenamefont {Cherukulappurath}, \citenamefont {Juanola~Parramon},
  \citenamefont {{Molina-Terriza}},\ and\ \citenamefont
  {Quidant}}]{volpeControllingOpticalField2009}%
  \BibitemOpen
  \bibfield  {author} {\bibinfo {author} {\bibfnamefont {G.}~\bibnamefont
  {Volpe}}, \bibinfo {author} {\bibfnamefont {S.}~\bibnamefont
  {Cherukulappurath}}, \bibinfo {author} {\bibfnamefont {R.}~\bibnamefont
  {Juanola~Parramon}}, \bibinfo {author} {\bibfnamefont {G.}~\bibnamefont
  {{Molina-Terriza}}}, \ and\ \bibinfo {author} {\bibfnamefont
  {R.}~\bibnamefont {Quidant}},\ }\href {\doibase 10.1021/nl901821s} {\bibfield
   {journal} {\bibinfo  {journal} {Nano Letters}\ }\textbf {\bibinfo {volume}
  {9}},\ \bibinfo {pages} {3608} (\bibinfo {year} {2009})}\BibitemShut
  {NoStop}%
\bibitem [{\citenamefont {Wo{\'z}niak}\ \emph {et~al.}(2015)\citenamefont
  {Wo{\'z}niak}, \citenamefont {Banzer},\ and\ \citenamefont
  {Leuchs}}]{wozniakSelectiveSwitchingIndividual2015}%
  \BibitemOpen
  \bibfield  {author} {\bibinfo {author} {\bibfnamefont {P.}~\bibnamefont
  {Wo{\'z}niak}}, \bibinfo {author} {\bibfnamefont {P.}~\bibnamefont {Banzer}},
  \ and\ \bibinfo {author} {\bibfnamefont {G.}~\bibnamefont {Leuchs}},\ }\href
  {\doibase 10.1002/lpor.201400188} {\bibfield  {journal} {\bibinfo  {journal}
  {Laser \& Photonics Reviews}\ }\textbf {\bibinfo {volume} {9}},\ \bibinfo
  {pages} {231} (\bibinfo {year} {2015})}\BibitemShut {NoStop}%
\bibitem [{\citenamefont {Guzzinati}\ \emph {et~al.}(2017)\citenamefont
  {Guzzinati}, \citenamefont {B{\'e}ch{\'e}}, \citenamefont {{Louren{\c
  c}o-Martins}}, \citenamefont {Martin}, \citenamefont {Kociak},\ and\
  \citenamefont {Verbeeck}}]{guzzinatiProbingSymmetryPotential2017}%
  \BibitemOpen
  \bibfield  {author} {\bibinfo {author} {\bibfnamefont {G.}~\bibnamefont
  {Guzzinati}}, \bibinfo {author} {\bibfnamefont {A.}~\bibnamefont
  {B{\'e}ch{\'e}}}, \bibinfo {author} {\bibfnamefont {H.}~\bibnamefont
  {{Louren{\c c}o-Martins}}}, \bibinfo {author} {\bibfnamefont
  {J.}~\bibnamefont {Martin}}, \bibinfo {author} {\bibfnamefont
  {M.}~\bibnamefont {Kociak}}, \ and\ \bibinfo {author} {\bibfnamefont
  {J.}~\bibnamefont {Verbeeck}},\ }\href {\doibase 10.1038/ncomms14999}
  {\bibfield  {journal} {\bibinfo  {journal} {Nature Communications}\ }\textbf
  {\bibinfo {volume} {8}},\ \bibinfo {pages} {14999} (\bibinfo {year}
  {2017})}\BibitemShut {NoStop}%
\bibitem [{\citenamefont {Alexander}\ \emph {et~al.}(2021)\citenamefont
  {Alexander}, \citenamefont {Flauraud},\ and\ \citenamefont
  {{Demming-Janssen}}}]{alexanderNearFieldMappingPhotonic2021}%
  \BibitemOpen
  \bibfield  {author} {\bibinfo {author} {\bibfnamefont {D.~T.~L.}\
  \bibnamefont {Alexander}}, \bibinfo {author} {\bibfnamefont {V.}~\bibnamefont
  {Flauraud}}, \ and\ \bibinfo {author} {\bibfnamefont {F.}~\bibnamefont
  {{Demming-Janssen}}},\ }\href {\doibase 10.1021/acsnano.1c06065} {\bibfield
  {journal} {\bibinfo  {journal} {ACS Nano}\ }\textbf {\bibinfo {volume}
  {15}},\ \bibinfo {pages} {16501} (\bibinfo {year} {2021})}\BibitemShut
  {NoStop}%
\bibitem [{\citenamefont {Wiecha}\ and\ \citenamefont
  {Muskens}(2020)}]{wiechaDeepLearningMeets2020}%
  \BibitemOpen
  \bibfield  {author} {\bibinfo {author} {\bibfnamefont {P.~R.}\ \bibnamefont
  {Wiecha}}\ and\ \bibinfo {author} {\bibfnamefont {O.~L.}\ \bibnamefont
  {Muskens}},\ }\href {\doibase 10.1021/acs.nanolett.9b03971} {\bibfield
  {journal} {\bibinfo  {journal} {Nano Letters}\ }\textbf {\bibinfo {volume}
  {20}},\ \bibinfo {pages} {329} (\bibinfo {year} {2020})},\ \Eprint
  {http://arxiv.org/abs/1909.12056} {arXiv:1909.12056} \BibitemShut {NoStop}%
\bibitem [{\citenamefont {An}\ \emph {et~al.}(2021)\citenamefont {An},
  \citenamefont {Zheng}, \citenamefont {Shalaginov}, \citenamefont {Tang},
  \citenamefont {Li}, \citenamefont {Zhou}, \citenamefont {Dong}, \citenamefont
  {Haerinia}, \citenamefont {Agarwal}, \citenamefont {{Rivero-Baleine}},
  \citenamefont {Kang}, \citenamefont {Richardson}, \citenamefont {Gu},
  \citenamefont {Hu}, \citenamefont {Fowler},\ and\ \citenamefont
  {Zhang}}]{anDeepConvolutionalNeural2021}%
  \BibitemOpen
  \bibfield  {author} {\bibinfo {author} {\bibfnamefont {S.}~\bibnamefont
  {An}}, \bibinfo {author} {\bibfnamefont {B.}~\bibnamefont {Zheng}}, \bibinfo
  {author} {\bibfnamefont {M.~Y.}\ \bibnamefont {Shalaginov}}, \bibinfo
  {author} {\bibfnamefont {H.}~\bibnamefont {Tang}}, \bibinfo {author}
  {\bibfnamefont {H.}~\bibnamefont {Li}}, \bibinfo {author} {\bibfnamefont
  {L.}~\bibnamefont {Zhou}}, \bibinfo {author} {\bibfnamefont {Y.}~\bibnamefont
  {Dong}}, \bibinfo {author} {\bibfnamefont {M.}~\bibnamefont {Haerinia}},
  \bibinfo {author} {\bibfnamefont {A.~M.}\ \bibnamefont {Agarwal}}, \bibinfo
  {author} {\bibfnamefont {C.}~\bibnamefont {{Rivero-Baleine}}}, \bibinfo
  {author} {\bibfnamefont {M.}~\bibnamefont {Kang}}, \bibinfo {author}
  {\bibfnamefont {K.~A.}\ \bibnamefont {Richardson}}, \bibinfo {author}
  {\bibfnamefont {T.}~\bibnamefont {Gu}}, \bibinfo {author} {\bibfnamefont
  {J.}~\bibnamefont {Hu}}, \bibinfo {author} {\bibfnamefont {C.}~\bibnamefont
  {Fowler}}, \ and\ \bibinfo {author} {\bibfnamefont {H.}~\bibnamefont
  {Zhang}},\ }\href {\doibase 10.1002/adom.202102113} {\bibfield  {journal}
  {\bibinfo  {journal} {Advanced Optical Materials}\ ,\ \bibinfo {pages}
  {2102113}} (\bibinfo {year} {2021})},\ \Eprint
  {http://arxiv.org/abs/2102.01761} {arXiv:2102.01761} \BibitemShut {NoStop}%
\bibitem [{\citenamefont {Wiecha}\ \emph {et~al.}(2021)\citenamefont {Wiecha},
  \citenamefont {Arbouet}, \citenamefont {Girard},\ and\ \citenamefont
  {Muskens}}]{wiechaDeepLearningNanophotonics2021}%
  \BibitemOpen
  \bibfield  {author} {\bibinfo {author} {\bibfnamefont {P.~R.}\ \bibnamefont
  {Wiecha}}, \bibinfo {author} {\bibfnamefont {A.}~\bibnamefont {Arbouet}},
  \bibinfo {author} {\bibfnamefont {C.}~\bibnamefont {Girard}}, \ and\ \bibinfo
  {author} {\bibfnamefont {O.~L.}\ \bibnamefont {Muskens}},\ }\href {\doibase
  10.1364/PRJ.415960} {\bibfield  {journal} {\bibinfo  {journal} {Photonics
  Research}\ }\textbf {\bibinfo {volume} {9}},\ \bibinfo {pages} {B182}
  (\bibinfo {year} {2021})},\ \Eprint {http://arxiv.org/abs/2011.12603}
  {arXiv:2011.12603} \BibitemShut {NoStop}%
\bibitem [{\citenamefont {Majorel}\ \emph {et~al.}(2022)\citenamefont
  {Majorel}, \citenamefont {Girard}, \citenamefont {Arbouet}, \citenamefont
  {Muskens},\ and\ \citenamefont {Wiecha}}]{majorelDeepLearningEnabled2022}%
  \BibitemOpen
  \bibfield  {author} {\bibinfo {author} {\bibfnamefont {C.}~\bibnamefont
  {Majorel}}, \bibinfo {author} {\bibfnamefont {C.}~\bibnamefont {Girard}},
  \bibinfo {author} {\bibfnamefont {A.}~\bibnamefont {Arbouet}}, \bibinfo
  {author} {\bibfnamefont {O.~L.}\ \bibnamefont {Muskens}}, \ and\ \bibinfo
  {author} {\bibfnamefont {P.~R.}\ \bibnamefont {Wiecha}},\ }\href {\doibase
  10.1021/acsphotonics.1c01556} {\bibfield  {journal} {\bibinfo  {journal} {ACS
  Photonics}\ }\textbf {\bibinfo {volume} {9}},\ \bibinfo {pages} {575}
  (\bibinfo {year} {2022})},\ \Eprint {http://arxiv.org/abs/2110.02109}
  {arXiv:2110.02109} \BibitemShut {NoStop}%
\end{thebibliography}
%

\end{document}